\documentclass[10pt,journal,compsoc]{IEEEtran}
\ifCLASSOPTIONcompsoc
  \usepackage[nocompress]{cite}
\else
  \usepackage{cite}
\fi
\ifCLASSINFOpdf
\else
\fi
\usepackage{amsfonts}    % Enable this line and disable the
\usepackage{color}
\usepackage{graphicx}
\usepackage[dvips]{epsfig}
\usepackage{graphics}
\usepackage{cite}
\usepackage{arydshln}
\usepackage{amsmath} % assumes amsmath package installed
\usepackage{amssymb}  % assumes amsmath package installed
\usepackage{subfigure}
\usepackage{subfloat}
\usepackage{caption}
\usepackage{algorithm} %ctan.org\pkg\algorithms
\usepackage{algpseudocode}
\usepackage{verbatim}
\usepackage{arcs}

\def\bt{\bar{\theta}}

\def \qed {\hfill \vrule height6pt width 6pt depth 0pt}
\def\bee{\begin{equation}}
\def\ene{\end{equation}}
\def\been{\begin{equation*}}
\def\enen{\end{equation*}}
\def\beq{\begin{eqnarray}}
\def\enq{\end{eqnarray}}

\newtheorem{pro}{Proposition}[section]
\newtheorem{lem}{Lemma}[section]

\newtheorem{cor}{Corollary}[section]
\newtheorem{thm}{Theorem}[section]

\begin{document}
\title{Mobile Social Services with Network Externality: From Separate Pricing to Bundled Pricing}

\author{Xuehe~Wang,~\IEEEmembership{Member,~IEEE,}
        Lingjie~Duan,~\IEEEmembership{Senior~Member,~IEEE,}
        and~Junshan~Zhang,~\IEEEmembership{Fellow,~IEEE}% <-this % stops a space
\IEEEcompsocitemizethanks{\IEEEcompsocthanksitem X. Wang and L. Duan are with the Pillar of Engineering Systems and Design, Singapore University of Technology and Design, Singapore 487372 (e-mail: xuehe\_wang@sutd.edu.sg; lingjie\_duan@sutd.edu.sg).
\IEEEcompsocthanksitem J. Zhang is with the School of Electrical, Computer and Energy Engineering, Arizona State University, Tempe, AZ 85287, USA (email: junshan.zhang@asu.edu).
\IEEEcompsocthanksitem This work was supported by the Singapore Ministry of Education Academic Research Fund Tier 2 under Grant MOE2016-T2-1-173.}% <-this % stops an unwanted space
}

\IEEEtitleabstractindextext{%
\begin{abstract}

Today, many wireless device providers choose to sell devices bundled with complementary mobile social services, which exhibit strong positive network externality. Taking a reverse engineering approach, this paper aims to quantify the benefits of selling devices and
complementary services under the following three strategies: separate pricing, bundled pricing, and hybrid pricing (both the separate and bundled options are offered). A comprehensive comparison of the above three strategies is carried out for two popular service models, namely physical connectivity sharing and virtual content sharing, respectively. These two sharing service models are two
popular examples under the emerging trend of sharing economy. We first study the physical service model where the provider (e.g., FON) offers users customized WiFi devices for indoor Internet access, and allows service subscribers to physically access all device owners' WiFi when traveling. Observing that all device-owners contribute to the connectivity sharing, we show, via a Stackelberg game theoretic approach, that bundled pricing outperforms separate pricing as long as the total cost of device and service is reasonably low to stimulate network externality. Further, hybrid pricing strictly dominates bundled pricing thanks to the pricing flexibility to keep high marginal profit of device-selling. Next, we investigate the virtual sharing service model where the provider (e.g., Apple) sells devices and device-supported applications. Different from the connectivity service model, in this model service subscribers directly contribute to the virtual content sharing, and the network externality can be fairly strong. We prove that hybrid pricing degenerates to bundled pricing if the network externality degree is larger than the average device valuation, which is in stark contrast with the connectivity service model in which hybrid pricing always outperforms bundled pricing.

\end{abstract}

}

\maketitle

\IEEEdisplaynontitleabstractindextext

% \fi
\IEEEpeerreviewmaketitle

\section{Introduction}

\IEEEPARstart{N}{owadays}, mobile social services are becoming a main driver for a wireless device provider's revenue growth. Different from traditional social services, mobile social services are enabled by new wireless devices and allow users to share physical wireless connectivity or virtual content with each other. For example, FON company\footnote{https://fon.com} not only sells WiFi routers for mobile Internet access at home but also introduces a WiFi sharing service for global connectivity by aggregating users' WiFi connectivity. It offers members access to more than 35 million crowdsourced WiFi hotspots in 1000 cities at the end of 2016 \cite{fon35}. Similarly, Apple provides wearable iWatch devices and promotes new virtual applications (e.g., Activity and Health, for users' measuring, planning and sharing fitness and health activities). These new social applications exhibit strong positive network externality among users and helped the iWatch sales hit 12 million units in 2015 \cite{iwatch2016}. We shall consider two types of mobile social services, namely physical connectivity sharing and virtual content sharing service models. In the first service model, the physical sharing service's value to a user depends on the number of device-owners who contribute to the device-supported service, and a user at a time is either using indoor home device or outdoor service. In contrast, in the virtual sharing model, application/service users alone generate content and contribute to the service value, which is independent of device value. The above two sharing service models are two popular ones under the emerging trend of sharing economy. Indeed, as users are provided with a community platform to efficiently share connectivity or information with unprecedented ease, the potential of a sharing economy is huge. A recent report by Ericsson Consumer Lab indicates that as many as three-quarters of smartphone owners are interested in this sharing platform \cite{consumerlab10hot}.

FON, OpenSpark and Whisher are the pioneers in launching the physical sharing model, stimulating the wireless sharing economy among users and providing global WiFi connectivity beyond home coverage \cite{niyato2007wireless}. Such mobile social services among users are enabled by secure sharing devices (e.g., Fonera routers for FON's case) and require central coordination and management \cite{afrasiabi2012pricing}. For example, to become a FON member, one needs to purchase a customized WiFi router called Fonera, which enables users to share WiFi bandwidth and access keys with each other. Other companies such as OpenSpark \cite{openspark} and Whisher \cite{whisher} follow a similar sharing business by including residential and even portable WiFi routers. Many of these providers including FON have chosen to use bundled pricing to sell the device and complementary social service as a package \cite{rafiei2013product}. This is intriguing and has motivated us to ask the first key question in this paper: why does a profit-maximizing provider prefer bundled pricing over separate pricing when providing the physical sharing service, and how is the pricing choice affected by the device and service costs?

In the second service model, many new wearable devices, such as iWatch and Samsung Gear, support Health and Fitness application services that enable users to easily monitor heart rate or track physical exercises in real time \cite{lee2016analysis}. By generating and sharing massive content (e.g., heart beats, fitness milestones and locations), users build strong social ties with each other. Unlike the physical sharing model, in this context users are flexible to use the device and virtual applications at any time, and the service users directly contribute to the social service's network externality. Moreover, the applications are no longer attractive to users without owning wearable devices and separate pricing does not work in this model. Given the two options of bundled pricing and hybrid pricing, we observe that the provider always prefers to bundle popular service applications (e.g., Activity app and Fitness app in iWatch) with devices; whereas some less popular services (e.g., Runtastic, Golfshot Pro and Golfplan in iWatch) are not necessarily bundled and users can buy devices only. This motivates us to ask the second key question: under which conditions should the provider bundle the content sharing service with device selling and skip providing device-only option?

To answer these two key questions, we study the optimal pricing choices for both service models. Our main contributions are summarized as follows.
\begin{itemize}
  \item \emph{A quantitative analysis of bundled pricing for device-supported mobile social services}: To our best knowledge, this is the first paper to study the bundled pricing for emerging wireless devices and complementary mobile social services with positive network externality. In Section \ref{sec_systemmodel}, we formulate the interactions between provider and heterogeneous users as a Stackelberg game: In stage I, the provider determines to adopt which pricing strategy (separate, bundled or hybrid pricing); In stage II, a user decides which product to buy based on his utilities, while taking into consideration the other users' decisions due to positive network externality.

  \item \emph{Separate versus bundled/hybrid pricing for connectivity sharing services}: In Section \ref{sec_trafficsharing}, we analyze provider's optimal pricing strategy by considering device-owners' contribution to the physical sharing service. It is shown that bundled pricing outperforms separate pricing as long as the total device and service cost is reasonably low to stimulate network externality. Hybrid pricing strictly dominates bundled pricing thanks to the pricing flexibility to keep high marginal profit of device-selling. Moreover, we show that the service cost is more critical than the device cost to affect the pricing choice.

  \item \emph{Bundled versus hybrid pricing for content sharing services}: In Section \ref{sec_contentsharing}, we analyze the provider's optimal pricing strategy by comparing the bundled pricing against the more general hybrid pricing, where service users directly contribute to the service value. Unlike the physical connectivity sharing model with comparable device and service values (for indoor and outdoor usages), here the service value is independent of device value and can be fairly large. Different from the connectivity service model in which hybrid pricing always outperforms bundled pricing, the provider is more likely to sell the device and service as a bundle as the degree of network externality $\lambda$ increases. To be specific, hybrid pricing degenerates to bundled pricing once the network externality degree is larger than the average device valuation.%, as the bundle product stimulates maximum network externality and service-selling profit
\end{itemize}

We have a few words on related work, most of which focuses on pricing of either wireless devices or services without considering the bundling, not to mention the network externality for mobile social services (e.g., \cite{afrasiabi2012pricing}, \cite{gizelis2011survey}, \cite{duan2013economics}). The work \cite{feldman2013pricing} discusses how to decide the price for a single type of good allocation (e.g., snow blower) which not only benefits the buyers allocated with the public good but also their neighbors under network externalities. The work \cite{munagala2014value} further studied the auction design for provisioning the single type of good under incomplete information and network externalities. Different from these two works, our study here focuses on two related types of goods (device and the device-supported service) for bundled pricing. While in recent economics models which consider bundled pricing or network externality (e.g., \cite{prasad2010optimal}, \cite{yan2011profit}, \cite{cohen2011truth}, \cite{wu2014exploring}, \cite{gong2016jsac}), their results for independent products cannot apply to our device-supported mobile social service models. The works \cite{niyato2008competitive}, \cite{duan2012duopoly}, \cite{duan2015pricing} study the pricing design under multi-providers' competition. Note that they only study a single type of good's pricing instead of two related types of goods with network externality in our paper. Still, we can imagine some intuitive results of pricing competition hold for our device-supported mobile social service model. For example, providers may differentiate in their pricing choices to avoid severe competition. As users' subscription decides the network coverage, we also expect providers to differentiate in their equilibrium coverage for provisioning different QoS. We plan to study multi-provider bundled pricing formally in the future. In this paper, we study the optimal pricing schemes for both connectivity and content sharing service models in the spirit of sharing economy, in which the device-supported mobile social services exhibit network externality. We compare the provider's optimal profits, under three pricing strategies, namely separate pricing, bundled pricing and hybrid pricing, and characterize the conditions under which we can take advantages of service bundling.

%they don't consider device-supported services and many results are based on simulations as analysis is complicated, which can't be directly applied to our models (e.g., \cite{prasad2010optimal}). Unlike the previous work,

\section{System Model}\label{sec_systemmodel}

Consider a monopoly provider who sells devices and complementary mobile social services to users, where users are heterogeneous in their valuations for device and service. The user population is normalized to be one unit. Due to the network externality, the quality of service (QoS) $Q(D_s)$ increases with the number of users $D_s$ contributing to the network externality, where $D_s\in[0,1]$. More details on $D_s$ will be presented for the two different models later. A particular user's valuations for the device and the service are denoted as $R_1$ and $f(Q(D_s))$, respectively. Two different users' valuations for device or service may be different. %Note that $n_v$ can be the number of users subscribing to the service or contributing to the social service coverage.

In separate pricing, the provider sells the device and service separately, and the utility of a user for buying the device only is
\bee\label{equ_Ueinitial} U_1=R_1-p_1, \ene
and his utility for buying the service only is\footnote{In the physical sharing service, a user has his own phone or laptop to use the service only option. However, the content sharing service should be supported by a wearable device and service-only option is not attractive to users.}
\bee\label{equ_Usinitial} U_{2}=f(Q(D_s))-p_2, \ene
where $p_1$ and $p_2$ are the device and service prices, respectively.

The user will buy the device if $U_1\geq 0$ and the service if $U_2\geq 0$. Given users' valuation distribution and network externality, we aim to derive the users' demands $D_1$ and $D_2\in[0,1]$ for the device and service, respectively. The provider's total profit under separate pricing is
\bee\label{equ_pi_s} \Pi_s(p_1,p_2)=(p_1-c_1)D_1+(p_2-c_2)D_2, \ene
where $c_1$ and $c_2$ are the unit costs of the device and service, respectively.\footnote{Our cost model can be extended to further include the one-time cost (e.g., for server deployment) for setting up the social service infrastructure. As long as the profit of service justifies the service cost, the service is finally provided and the one-time cost does not affect the pricing.}

In bundled pricing, the provider sells the device and service as a bundle, and the subscribing user's utility is
\bee\label{equ_bundle} U_{12}=R_1+f(Q(D_{s}))-p_{12}. \ene
where $p_{12}$ is the bundled price.
 %Note that under bundle pricing, $n_b=n_s$.

Users will buy the bundle product if the utility $U_{12}\geq 0$.
%By collecting the bundle fees for the device and service from the users,
It follows that the provider's total profit under bundled pricing is
\bee\label{equ_pi_b} \Pi_b(p_{12})=(p_{12}-c_1-c_2)D_{12}, \ene
where $D_{12}\in[0,1]$ is the demand for the bundle product.

More generally, the provider can choose hybrid pricing by providing both the separate and bundled options. For example, if the provider offers device-only and bundled products to users with prices $p_1$ and $p_{12}$, the provider's total profit under hybrid pricing changes from (\ref{equ_pi_s}) and (\ref{equ_pi_b}) to
\bee\label{equ_Pimixbundle} \Pi_h(p_1,p_{12})=(p_1-c_1)D_1+(p_{12}-c_1-c_2)D_{12}. \ene

\begin{figure}
\centering\includegraphics[scale=0.35]{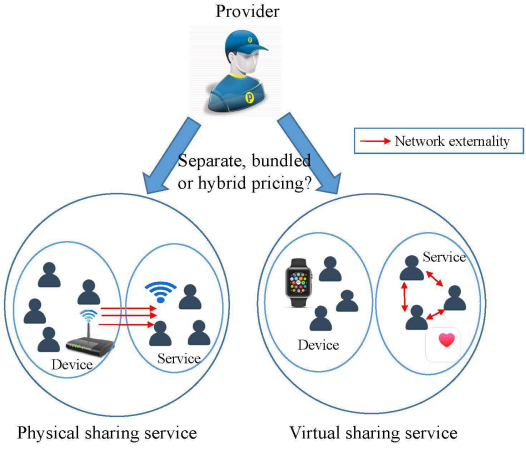}\caption{System model of the two service models}\label{sysmodel}
\end{figure}

Next, we specify the device value $R_1$ and the social value $f(Q(D_s))$ for two different device-supported mobile social service models, as illustrated in Fig. \ref{sysmodel}.

\begin{itemize}
  \item In the physical sharing service model, heterogeneous users split their time at home and outside differently. We denote a user's mobility factor as $\alpha\leq 1$, and $1-\alpha$ means his time proportion at home. Here, $\alpha$ is different for different users corresponding to their different mobility patterns or habits. As all purchased devices (e.g., Fonera routers for FON's case) contribute to the network externality by default, $D_s$ is the number of device buyers (e.g., $D_1$ in separate pricing and $D_{12}$ in bundled pricing) and $Q(D_s)$ represents the outdoor WiFi coverage by aggregating all devices' connectivity. Thus, we reasonably assume $R_1=1-\alpha$ for Internet access benefit at home and $f(Q(D_s))=\alpha Q(D_s)$ for outdoor Internet access benefit, depending on both mobility and the outdoor WiFi coverage. The device and service values are comparable and we bound both values by value 1.

  \item In the virtual sharing service model, it is service subscribers instead of all device owners that contribute to the network externality, and users are flexible to use the device and service at any time. A user's time occupancies in the device and service are no longer complementary. A particular user's valuation for the device is denoted as $R_1\in [0,\bar{\theta}]$, where $\bar{\theta}$ tells the maximum possible value one associates to the local device usage (e.g., time watching, calendar management, and web browsing). Different from the physical sharing service model, in this case service users generate and share content themselves and $D_s$ is the number of users subscribing to the social service. We assume a general valuation model for the service including an intrinsic service value $R_2\in [0,1]$ (e.g., heart beat monitoring in Health app) and an external social value $\lambda Q(D_s)$ (e.g., sharing heart beats in the same app), where $\lambda$ is the degree of network externality that is higher for popular sharing service. The total service value to a user is $f(Q(D_s))=R_2+\lambda Q(D_s)$. Note that in the physical sharing model, a user gets access to at most one WiFi hotspot at home or outside at a time and the physical connectivity sharing service's value is bounded by $1$. However, the virtual sharing service's value here is independent of device value and can be reasonably large, telling users' flexibility to generate and share massive content with each other.
\end{itemize}

In both service models, we formulate the two-stage interactions between the provider and users as a Stackelberg game. In Stage I, the provider determines the pricing strategy and the corresponding price profile to maximize its profit. In Stage II, the users decides whether and which option to buy based on his device and service utilities as well as the other users' decisions due to network externality.
We will use backward induction to first analyze users' equilibrium purchase at Stage II for any given pricing strategy and then the provider's optimal pricing strategy at Stage I by taking into consideration the users' best responses.

%\newcounter{mytempeqncnt}
%\begin{figure*}[ht]
%\setcounter{mytempeqncnt}{\value{equation}}
%\setcounter{equation}{13}
%\bee\label{equ_p11}\begin{split} p_{1,1/2}=&\frac{19+4c_1}{24}+\frac{-1\pm\sqrt{3}i}{2}\sqrt[3]{\kappa+\sqrt{\kappa^2+\Big(\frac{14+8c_1}{24}-\frac{(19+4c_1)^2}{576}\Big)^3}}\\
%&+\frac{-1\mp\sqrt{3}i}{2}\sqrt[3]{\kappa-\sqrt{\kappa^2+\Big(\frac{14+8c_1}{24}-\frac{(19+4c_1)^2}{576}\Big)^3}}, \end{split}\ene
%
%with \been\begin{split} \kappa=-\frac{(19+4c_1)(14+8c_1)}{384}+\Big(\frac{19+4c_1}{24}\Big)^3+\frac{c_2^2+4c_1+3}{16}, \end{split}\enen
%\setcounter{equation}{\value{mytempeqncnt}}
%\hrulefill
%\end{figure*}

In the following, we will analyze the optimal pricing choice for the physical sharing model in Section \ref{sec_trafficsharing} and then for virtual sharing model in Section \ref{sec_contentsharing}. The generalization of our models and results is discussed in Section \ref{sec_extends}. %\emph{Most of the proofs are quite lengthy and are given in our online technical report \cite{world1973technical}.}

\section{Optimal Pricing for Physical Connectivity Sharing Services}\label{sec_trafficsharing}
%For ease of exposition, \footnote{In practice, $Q(D_s)$ may be less than 1 (e.g., $Q(D_s)=0.9D_s$) depending how users' home distribution covers the area spatially, and our analysis and results can be extended to this situation. Intuitively, we just expect fewer users subscribing to the service and smaller network externality.}

In the physical sharing model, service quality $Q(D_s)$ is intimately related to the total WiFi coverage and increases with the number of device buyers. Generally, service quality $Q(D_s)$ (in term of total service coverage) increases quickly when $D_s$ is small and slowly when $D_s$ is large. This is because the access points can be densely distributed for large $D_s$, which results in overlap WiFi coverage \cite{manshaei2008wireless}. Thus, we set the service quality as
\bee\label{equ_QoSgamma} Q(D_s)=D_s^{\gamma}, \ene
where $\gamma\leq 1$ measures the service quality sensitivity. The small $\gamma$ captures the case where the service quality increases very quickly at the beginning and then very slowly. For analysis tractability, we consider the case of $\gamma = 1$ and $Q(D_s)=1$ tells a full coverage through the WiFi sharing. Later in Section \ref{sec_connectivity_hybrid} we will also investigate the general $\gamma$ distribution using simulations. For a user with mobility factor $\alpha\in[0,1]$, $1-\alpha$ is his local benefit for using home device only. By subscribing to the social service, the user receives the service benefit $\alpha D_s$. The mobility factor $\alpha\in[0,1]$ is assumed to follow uniform distribution. For example, a housewife who stays home mostly perceives $\alpha=0$, while a salesman who travels frequently perceives $\alpha=1$ as in \cite{duan2015pricing}. Our results can be extended to normal distribution for user mobility without major change of engineering insights (see
Section \ref{sec_normal_connectivity} for details). In the following, we will analyze users' behaviors and the provider's profits under separate pricing, bundled pricing and hybrid pricing, respectively.

\subsection{Separate Pricing for Connectivity Sharing Services}\label{sec_wifisep}

Under separate pricing, the provider sells the device and service separately. Given separate prices $p_1$ and $p_2$, for a user with mobility factor $\alpha$, the utility function for buying the device is
\bee\label{equ_Ue1} U_1(\alpha)=1-\alpha-p_1, \ene
which is derived from (\ref{equ_Ueinitial}) by letting $R_1=1-\alpha$.

His utility for buying the service is
\bee\label{equ_Us} U_2(\alpha)=\alpha D_1-p_2, \ene
which derived from (\ref{equ_Usinitial}) by letting $f(Q(D_s))=\alpha D_1$. As $D_s$ is the number of device buyers, we have $D_s=D_1$, which is not given but depends on all users' decisions on buying the device. The separate pricing should satisfy $0\leq p_1,p_2\leq 1$. Otherwise, all users receive negative payoffs and will not subscribe to device or service.

In the following, we use backward induction to first analyze the users' equilibrium decisions for any given separate pricing ($p_1, p_2$) in Stage II, and then analyze the provider's optimal separate pricing design in Stage I.

%Observe that for Stage II, it is nontrivial to derive users' subscription distribution, as a user's decision is affected by the others' and $D_1$ according to (\ref{equ_Us}). In the following, given any $D_1$, we first examine users' best response and then determine the equilibrium device demand $D_1^*$ as well as service demand $D_2^*$.

\begin{lem}\label{lem_wifialpha} Users of low mobility will buy the device for intense home WiFi usage, whereas those with high mobility will buy the service for intense outdoor usage. In particular, the equilibrium demand for the device is
\bee\label{equ_sep_D1*} D_1^*(p_1)=1-p_1, \ene
and the equilibrium demand for the service is
\bee\label{equ_sep_D2*} D_2^*(p_1, p_2)=1-\frac{p_2}{1-p_1}. \ene
%which is decreasing in both $p_1$ and $p_2$.
\end{lem}

\textbf{Proof:} It is clear that users with $U_1(\alpha)\geq 0$ will purchase the device, telling that $\alpha\leq 1-p_1$. According to (\ref{equ_Us}), given any $D_1$, a user will buy the complementary service if $\alpha\geq\frac{p_2}{D_1}$. Since $\alpha\in[0,1]$ follows uniform distribution, we have $D_1^*(p_1)=1-p_1$, and users with $\alpha\geq\frac{p_2}{1-p_1}$ will buy the service, i.e., $D_2^*(p_1, p_2)=1-\frac{p_2}{1-p_1}$.  \qed

Note that if $\gamma$ in (\ref{equ_QoSgamma}) is less than $1$, the equilibrium demand for the device is the same as (\ref{equ_sep_D1*}) but the equilibrium demand for the service changes from (\ref{equ_sep_D2*}) to $D_2^*(p_1, p_2)=1-\frac{p_2}{(1-p_1)^{\gamma}}$.

Equation (\ref{equ_sep_D2*}) tells how the device buyer population contributes to the service value. As $p_1$ increases, the number of device-owners $D_1^*$ decreases in (\ref{equ_sep_D1*}) and the the aggregated WiFi coverage or service quality decreases. Thus we expect a smaller demand $D_2^*$ in the service. We also notice that $D_2^*\geq 0$ only if $p_1+p_2\leq 1$.
After determining the users' equilibrium demands $D_1^*$ and $D_2^*$ in Stage II, we are ready to study Stage I by choosing optimal ($p_1, p_2$) to maximize the profit, i.e.,
\bee\label{equ_pro} \max_{0\leq p_1,p_2\leq 1} \Pi_s(p_1, p_2)=(p_1-c_1)D_1^*(p_1)+(p_2-c_2)D_2^*(p_1,p_{2}). \ene
%subject to $~~~~~~~~~~~~~~~0\leq p_1,p_2\leq 1.$
% $$p_1+p_2\leq 1.$$
%The provider's optimal separate pricing is summarized in the following proposition.

Note that $0\leq c_1,c_2\leq 1$ for the device and service costs, otherwise, the provider will not offer the device or service in the first place.
Problem (\ref{equ_pro}) is a non-convex optimal problem as the objective is not concave in $p_1$. Despite of this, we are still able to derive the optimal solutions in closed-form by examining the profit objective's relationship with $p_1$ and $p_2$ in different feasible regions.

\begin{pro}\label{pro_sep_wifi} Under the separate pricing strategy, the optimal device price $p_1^*$ and service price $p_2^*$ are determined as follows:
\begin{itemize}
\item If the social service is finally provided, the optimal device price $p_1^*$ is the minimal solution of
    \bee\label{equ_sep_solvep1} \frac{3}{4}-2p_1+c_1+\frac{c_2^2}{4(1-p_1)^2}=0. \ene
    The optimal service price $p_2^*$ is decreasing in $p_1^*$ and is given by
%\newcounter{mytempeqncnt1}
%\setcounter{mytempeqncnt1}{\value{equation}}
%\setcounter{equation}{13}
\bee\label{equ_sep_p2*} p_2^*=\frac{1+c_2-p_1^*}{2}. \ene
\item If the social service is not provided due to high costs, the optimal device price is \bee p_1^*=\frac{1+c_1}{2}.\ene
The resulting device-only profit is \bee \Pi_1(p_1^*)=\frac{(1-c_1)^2}{4}. \ene
\end{itemize}
\end{pro}

The proof of Proposition \ref{pro_sep_wifi} is given in Appendix \ref{app_pro_sep_wifi}. We have a few remarks on the relationship between $p_1^*$ and $p_2^*$ when the social service is provided. As the device price increases, fewer users will buy the device and contribute to the network externality of the service. This leads to a less valuable service and a lower service price.

To determine when the service is finally provided, we have the following corollary.

\begin{cor}\label{cor_sep_connectivity}
If $c_1+2c_2\leq 1$ with $0\leq c_1,c_2\leq 1$, the provider offers both device and service at the optimal separate pricing.
\end{cor}

According to Corollary \ref{cor_sep_connectivity}, we can see that both $c_1$ and $c_2$ affect the provision of service. As $c_1$ increases, $p_1^*$ increases and the device demand $D_1^*$ in (\ref{equ_sep_D1*}) decreases, leading to a smaller network externality to justify the service provision.

\begin{figure} \centering
\subfigure[$c_1=0.05, c_2=0.05$] { \label{alpha005005}
\includegraphics[width=0.6\columnwidth]{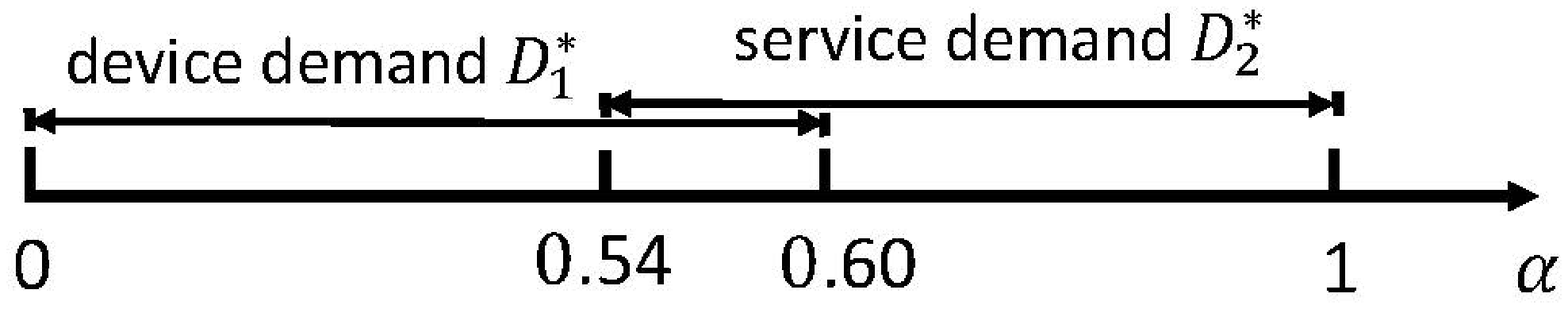}
}\\
\subfigure[$c_1=0.05, c_2=0.2$] { \label{alpha00502}
\includegraphics[width=0.6\columnwidth]{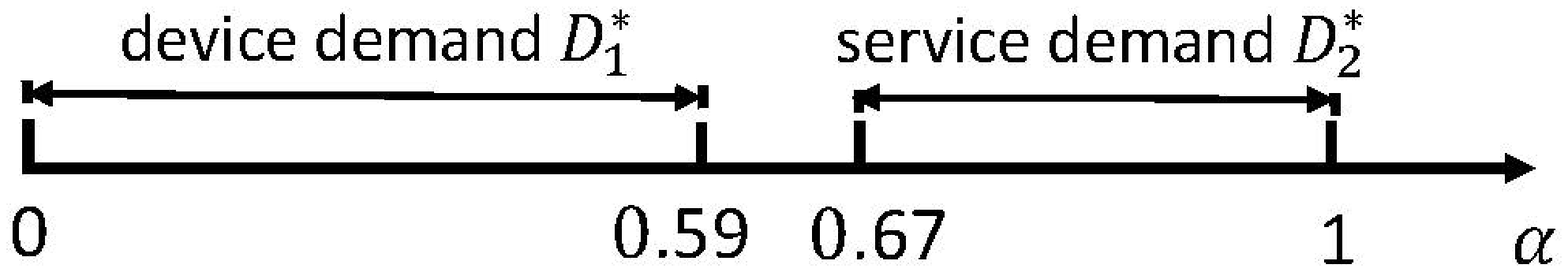}
}\\
\subfigure[$c_1=0.3, c_2=0.2$] { \label{alpha0302}
\includegraphics[width=0.6\columnwidth]{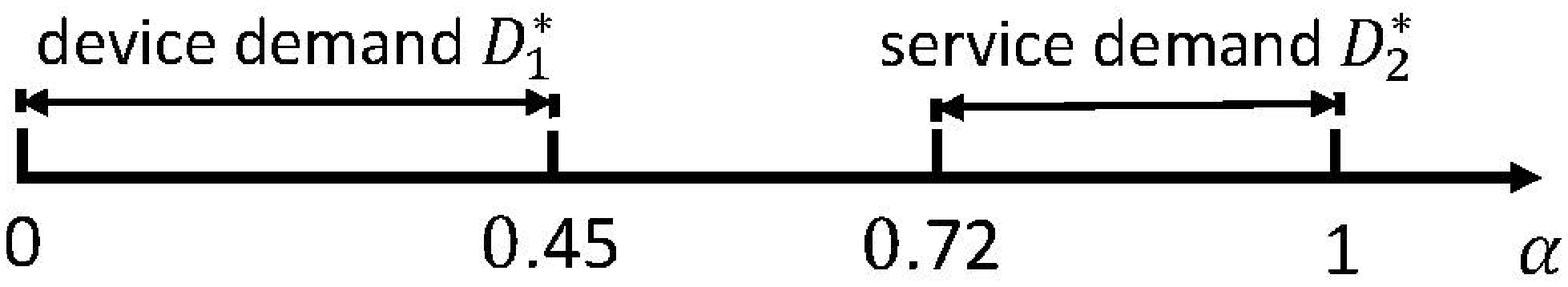}
}
\caption{Illustration of users' choices under optimal separate pricing $(p_1^*, p_2^*)$ when the social service is provided ($c_1+2c_2\leq 1$)}
\label{userchoice}
\end{figure}

After analyzing the whole two-stage game, as an illustrative example, Fig. \ref{userchoice} depicts how users' equilibrium choices are finally affected by costs $c_1$ and $c_2$ when the service is provided. In Fig. \ref{alpha005005}, $c_1, c_2$ are small, and some users buy both device and service due to affordable device and service prices. Fig. \ref{alpha00502} shows that not only the service demand $D_2^*$ decreases but also the device demand $D_1^*$ decreases as service cost $c_2$ increases. This is because that the service is less profitable and it is costly to maintain a high device demand for supplying high network externality of the service. As the device cost $c_1$ increases, Fig. \ref{alpha0302} shows service demand decreases as fewer device-owners under higher device price contribute to the service's value.

\subsection{Bundled Pricing for Connectivity Sharing Services and Pricing Comparison}\label{sec_bundle_wifi}

Under bundled pricing, the provider sells the device and service as a bundle. Given bundle price $p_{12}$, for a user with mobility factor $\alpha$, the utility function for buying the bundle package is
\bee\label{equ_Ubwifi} U_{12}(\alpha)=(1-\alpha)+\alpha D_{12}-p_{12}, \ene
which is derived from (\ref{equ_bundle}).
Here all subscribers buy the device (i.e., $D_s=D_{12}$). We require $p_{12}\leq 1$, otherwise, all users received negative payoffs and will not subscribe.
%Note that, under bundle pricing, $n_b=n_e=n_s$.

Equation (\ref{equ_Ubwifi}) tells that a user's decision is affected by the other users' decisions $D_{12}$. Letting $U_{12}(\alpha)\geq 0$, only users with $\alpha\leq\frac{1-p_{12}}{1-D_{12}}$ will subscribe to the service. Then we have
\bee\label{equ_n_b*} D_{12}=\frac{1-p_{12}}{1-D_{12}}, \ene
which has two roots. Note that if $\gamma<1$, the equilibrium demand changes from (\ref{equ_n_b*}) to $D_{12}=\frac{1-p_{12}}{1-D_{12}^{\gamma}}$.

The profit-maximizing provider prefers to choose the larger root as the equilibrium demand, i.e.,
\bee\label{equ_nbproof1} D_{12}^*(p_{12})=\frac{1+\sqrt{4p_{12}-3}}{2}. \ene
We will discuss how to make the users choose the larger root later.

Recall that in Stage I, the provider's objective is
\bee\label{equ_connectivity_bundle} \max_{0\leq p_{12}\leq 1} \Pi_b(p_{12})=(p_{12}-c_1-c_2)D_{12}^*. \ene

Note that $c_1+c_2\leq 1$, otherwise, the profit received by the provider is negative and the bundle product is not provided.
By solving problem (\ref{equ_connectivity_bundle}), the provider's optimal bundle pricing is as follows.

\begin{pro}\label{pro_wifi_bundle} Under the bundled pricing strategy, the optimal bundle price is the maximum $p_{12}^*=1$ and all users are still included ($D_{12}^*=1$). The resulting profit is
\begin{equation}\label{equ_connectivity_pib*} \Pi_b(p_{12}^*)=1-c_1-c_2. \end{equation}
\end{pro}

%\begin{figure}
%\centering\includegraphics[scale=0.29]{dynamicD12}
%\caption{Convergence of users' demand $D_{12}$ to the equilibrium $D_{12}^*=1$ given the initial demand $D_{12,t=0}=0.6$ under bundled pricing}\label{dynamicD}
%\end{figure}

The proof is given in Appendix \ref{app_pro_wifi_bundle}. %{\color{blue}{To ensure that the users will converge to the larger equilibrium demand (\ref{equ_nbproof1}) for benefit of obtaining a larger profit, the provider can first use an introductory price (e.g., a discount price $p_{12}^0=1-\varepsilon$, where $\varepsilon>0$ is infinitesimal) to let the initial demand $D_{12,t=0}$ achieve a value larger than $D_{12}^*(p_{12}^0)$ given in (\ref{equ_nbproof1}). Then, at time $t+1$, users update best responses according to $p_{12}^0$ and $D_{12,t}$, and the users' demand evolves according to
%\bee D_{12,t+1}=\frac{1-p_{12}^0}{1-D_{12,t}}=\frac{\varepsilon}{1-D_{12,t}}, \ene
%where the first equality is derived similar to (\ref{equ_n_b*}).
%
%Now we compare $D_{12,t}$ and $\frac{1-p_{12}^0}{1-D_{12,t}}$. Note that $D_{12,t}=\frac{1-p_{12}^0}{1-D_{12,t}}$ when $D_{12,t}=\frac{1\pm\sqrt{4p_{12}^0-3}}{2}$. Thus, it is easy to check that $D_{12,t}<\frac{1-p_{12}^0}{1-D_{12,t}}$ once $D_{12,t}>\frac{1+\sqrt{4p_{12}^0-3}}{2}$, which results in $D_{12,t+1}=\frac{1-p_{12}^0}{1-D_{12,t}}>D_{12,t}$. Therefore, by letting $D_{12,t=0}>\frac{1+\sqrt{4p_{12}^0-3}}{2}$, as shown in Fig. \ref{dynamicD}, users' demand $D_{12,t}$ increases over time $t$ and converge to the equilibrium demand $D_{12}^*=1$ in Proposition \ref{pro_wifi_bundle}. Then the provider can turn to maximum price $p_{12}^*=1$.}} %given a relatively large initial demand $D_{12}=0.6$, users' demand will converge to the equilibrium demand $D_{12}^*=1$ given in Proposition 3.2.This can be guaranteed by $D_{12}(0)>\frac{1+\sqrt{4p_{12}^0-3}}{2}$.

In reality, the unit cost of the service $c_2$ should not be very small due to the infrastructure and maintenance fee \cite{enck2011defending}. To rule out some trivial cases, in the following, we consider $c_2\geq 0.05$. By comparing the optimal profits under separate pricing and bundled pricing strategies in Propositions \ref{pro_sep_wifi} and \ref{pro_wifi_bundle}, we have the following result.

\begin{thm}\label{thm_wifi_compare} The provider prefers bundled pricing over separate pricing if
\bee\label{equ_thmwificompare} \frac{(1-c_1)(3+c_1)}{4}>c_2. \ene
Otherwise, the provider chooses separate pricing.
\end{thm}

%\textbf{Proof:} According to Propositions \ref{pro_sep_wifi} and \ref{pro_wifi_bundle}, under separate pricing, we have $p_1^*+p_2^*\leq p_{12}^*=1$. Since all users buy the bundle product and the bundle price $p_{12}^*$ reaches the maximum reasonable value $1$ under bundled pricing, thus we can verify that the bundle pricing always outperforms the separate pricing if the service is provided, i.e., $\Pi_s(p_1^*,p_2^*)<\Pi_b(p_{12}^*)$. Therefore, the separate pricing could be better than the bundle pricing if and only if the provider sell the device only. By comparing $\Pi_1(p_1^*)$ with $\Pi_b(p_{12}^*)$, (\ref{equ_thmwificompare}) is obtained. \qed

\begin{figure}
\centering\includegraphics[scale=0.5]{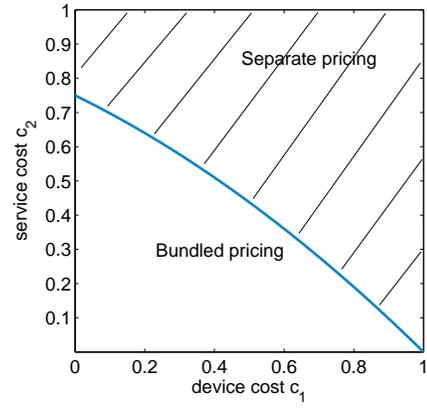}\caption{Comparison between bundled pricing and separate pricing for connectivity sharing service}\label{compare_connectivity}
\end{figure}

Notice that the left-hand-side of (\ref{equ_thmwificompare}) is decreasing in $c_1$. Fig. \ref{compare_connectivity} shows that if the total device and service cost is small, the bundled pricing outperforms the separate pricing as it stimulates strong network externality. If device cost $c_1$ is high, bundled pricing aiming to stimulate network externality has low marginal profit of device-selling. Thus, it is better to choose flexibly separate pricing to charge a high device price. If service cost $c_2$ is high, the marginal profit of service-selling is low and it is not worthwhile to sacrifice device-selling profit for network externality of the service. From Fig. \ref{compare_connectivity}, we can also see that the service cost $c_2$ is more sensitive than the device cost $c_1$ to decide the optimal pricing choice. Intuitively, as the service is device-supported, the bundled pricing sacrifices some device-selling profit and must be justified by high marginal profit of service-selling or low service cost.

\newcounter{myfig1}
\begin{figure*}[ht]
\setcounter{myfig1}{\value{figure}}
\setcounter{figure}{4}
\centering
\subfigure[High price regime $\bt+\lambda D_{12}<p_{12}<\bt+1+\lambda D_{12}$]{\label{pure1}
\begin{minipage}{.30\textwidth}
\includegraphics[width=1\textwidth]{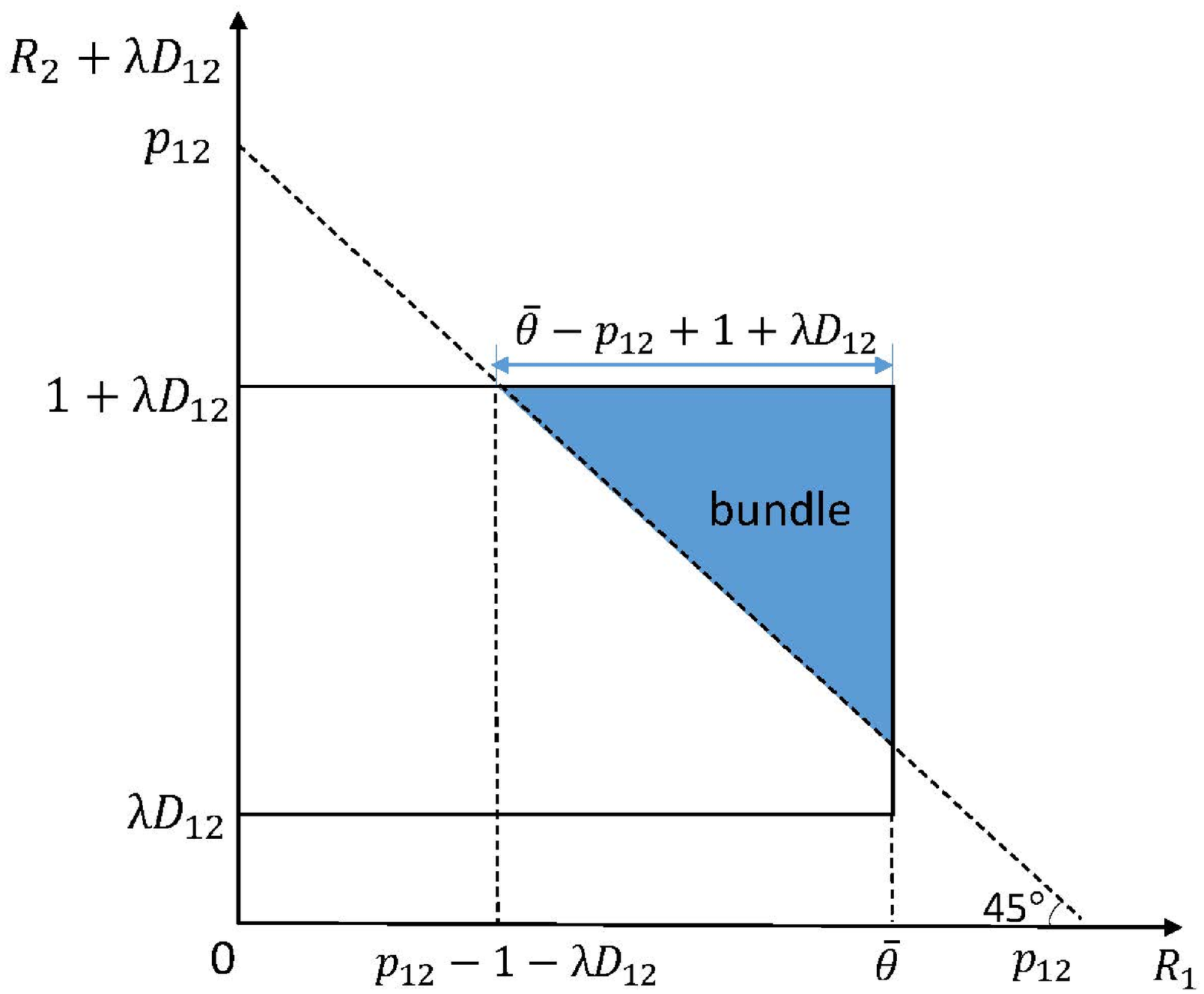}
\end{minipage}
}
\subfigure[Medium price regime $1+\lambda D_{12}<p_{12}<\bt+\lambda D_{12}$]{\label{pure2}
\begin{minipage}{.32\textwidth}
\includegraphics[width=1\textwidth]{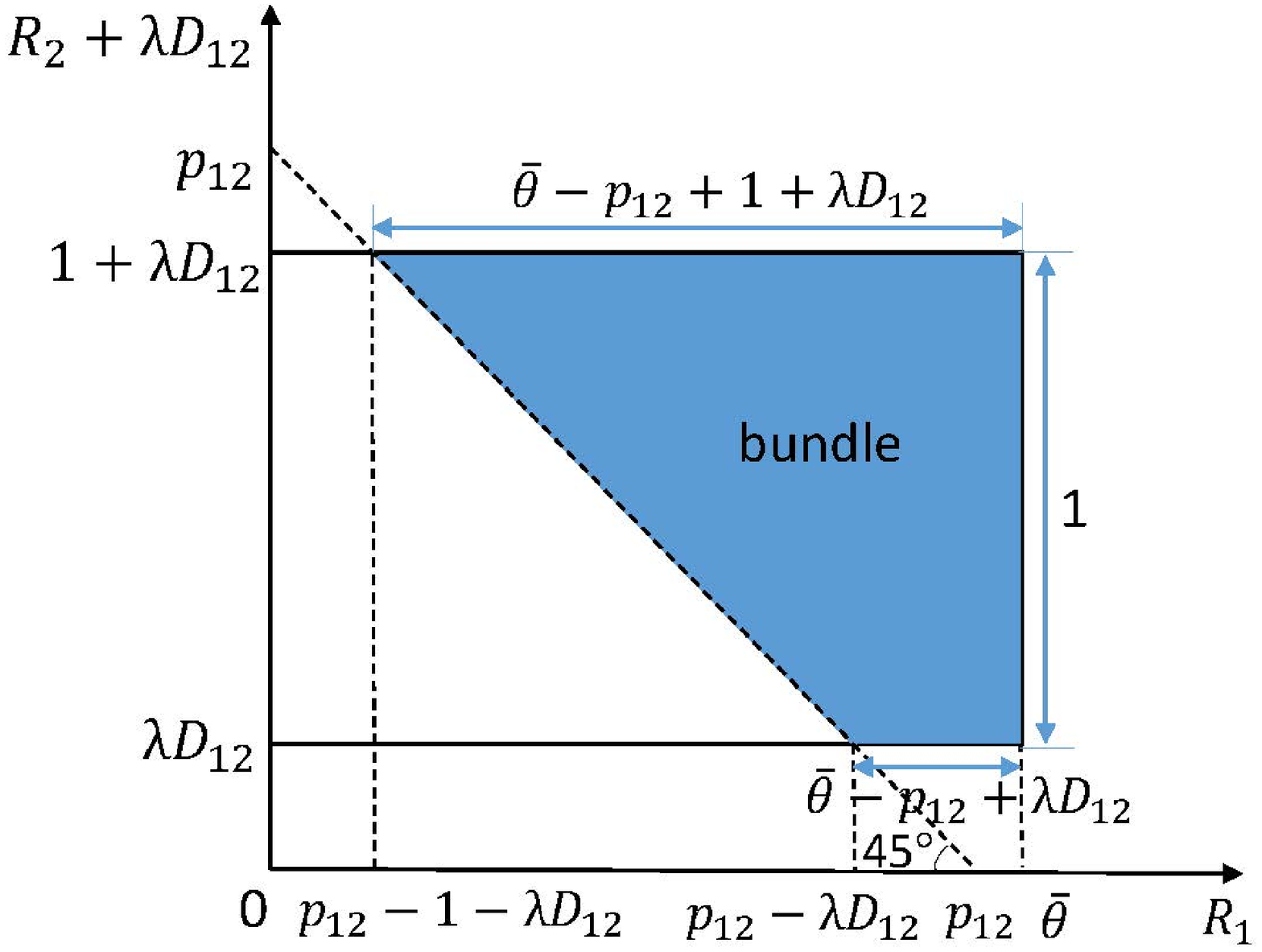}
\end{minipage}
}
\subfigure[Low price regime $\lambda D_{12}<p_{12}<1+\lambda D_{12}$]{\label{pure3}
\begin{minipage}{.32\textwidth}
\includegraphics[width=1\textwidth]{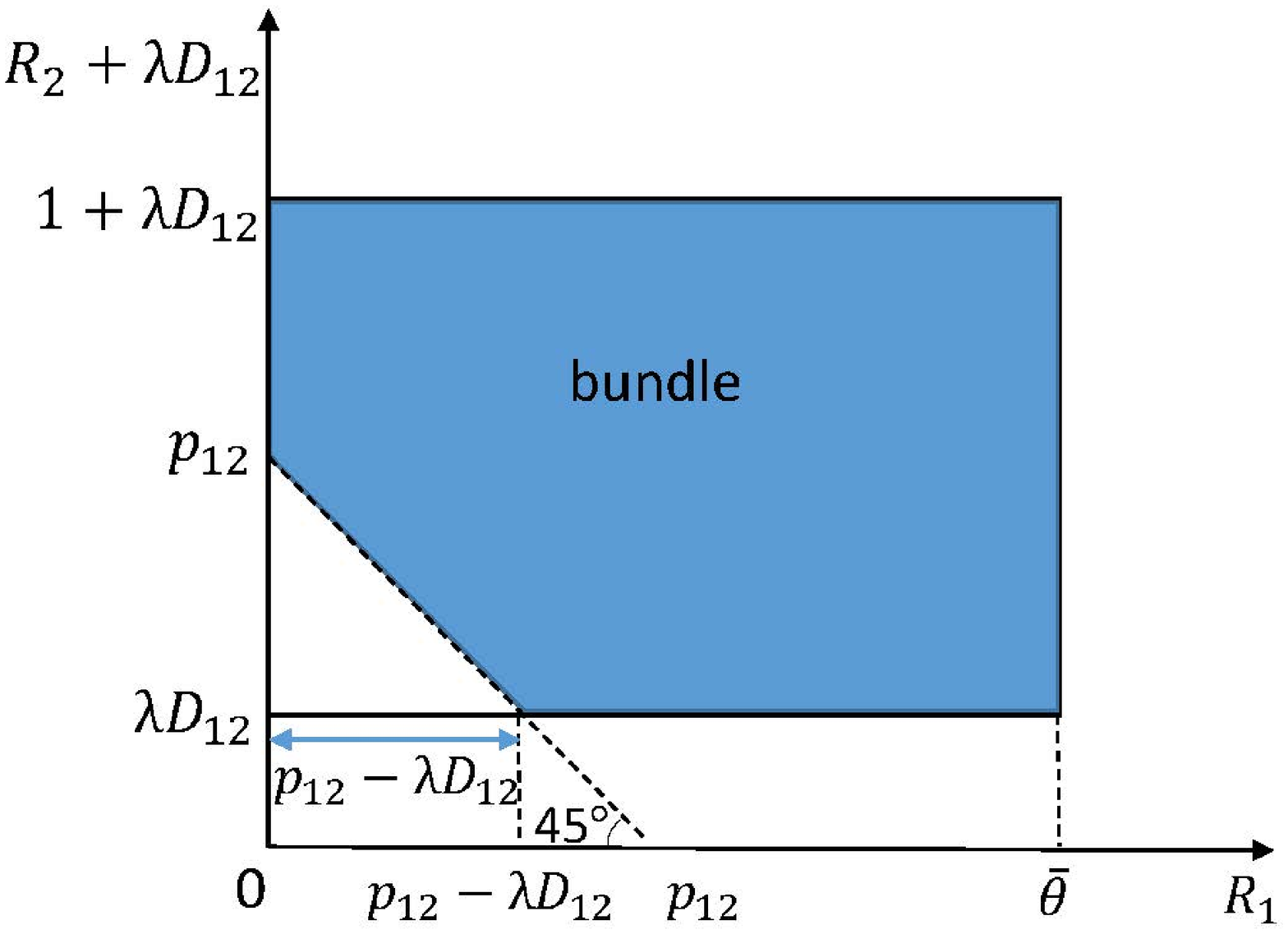}
\end{minipage}
}
\caption{Demand analysis for bundled pricing in the content sharing service model}\label{figBundled}
\setcounter{figure}{\value{myfig1}}
\end{figure*}

\subsection{Hybrid Pricing for Connectivity Sharing Services and Pricing Comparison}\label{sec_connectivity_hybrid}

As shown above, separate pricing has the flexibility to charge device and service differently, whereas bundled pricing can stimulate strong network externality for the device-supported service. It is natural to ask if there is a pricing strategy which can take advantages of both pricing strategies. In this subsection, we consider the case when the provider offers device-only and bundle options simultaneously. Users who purchase the device-only or bundle options contribute to the WiFi coverage, i.e., $D_s=D_1+D_{12}$.

Given the device prices $p_1\in[0,1]$ and bundle price $p_{12}\geq p_1$, the utility of a user for buying the device is given in (\ref{equ_Ue1}) and the utility for buying the bundle package is
\begin{equation} U_{12}(\alpha)=(1-\alpha)+\alpha (D_1+D_{12})-p_{12}, \end{equation}
which is equivalent to the case that the user will separately pay another price $p_{12}-p_1$ to obtain the service after paying the device.

\begin{lem}\label{lem_mixbundle} A user makes equilibrium decisions in Stage II according to his mobility factor $\alpha$:
\begin{itemize}
  \item If $0<\alpha\leq\min(1-p_1,\frac{p_{12}-p_1}{D_1+D_{12}})$, then the user will buy the device only.
  \item If $\frac{p_{12}-p_1}{D_1+D_{12}}\leq\alpha\leq\min(1, \frac{1-p_{12}}{1-D_1-D_{12}})$, then the user will buy the bundle to further benefit from the service.
\end{itemize}
\end{lem}

\textbf{Proof:} A user with mobility factor $\alpha\leq 1-p_1$ has $U_1\geq 0$ and will purchase the device. The user will further buy the social service if $U_{12}>U_1$ or simply $\alpha\geq\frac{p_{12}-p_1}{D_1+D_{12}}$. Then we conclude that the user with $0<\alpha\leq\min(1-p_1,\frac{p_{12}-p_1}{D_1+D_{12}})$ will buy the device only. When $\alpha\geq\frac{p_{12}-p_1}{D_1+D_{12}}$ and $U_{12}\geq0$, the user will buy the bundle. That is,$\frac{p_{12}-p_1}{D_1+D_{12}}\leq\alpha\leq\min(1, \frac{1-p_{12}}{1-D_1-D_{12}})$.\qed

It should be noted that some users of high mobility $\alpha$ and negative device-only benefit (i.e., $U_1(\alpha)<0$) may still subscribe to the bundle. This is because the social service benefit is large enough to justify the negative device-only utility.

According to Lemma \ref{lem_mixbundle}, the equilibrium demand $D_1^*$ for the device satisfies
\bee\label{equ_sep_mb_D1} D_1=\min(1-p_1,\frac{p_{12}-p_1}{D_1+D_{12}}), \ene
and the equilibrium demand $D_{12}^*$ for the bundle product satisfies
\bee\label{equ_sep_mb_D3} D_{12}=\min(1, \frac{1-p_{12}}{1-D_1-D_{12}})-\frac{p_{12}-p_1}{D_1+D_{12}}. \ene

Note that if $\gamma$ in (\ref{equ_QoSgamma}) is less than $1$, the equilibrium demand for device-only changes from (\ref{equ_sep_mb_D1}) to $D_1=\min(1-p_1,\frac{p_{12}-p_1}{(D_1+D_{12})^{\gamma}})$, and the equilibrium demand for the bundle product changes from (\ref{equ_sep_mb_D3}) to $D_{12}=\min(1, \frac{1-p_{12}}{1-(D_1+D_{12})^{\gamma}})-\frac{p_{12}-p_1}{(D_1+D_{12})^{\gamma}}$.

By solving (\ref{equ_sep_mb_D1}) and (\ref{equ_sep_mb_D3}) jointly, we can derive the equilibrium demands $D_1^*(p_1, p_{12})$ and $D_{12}^*(p_1, p_{12})$ as functions of $p_1$ and $p_{12}$. By substituting them back to analyze $\Pi_h(p_1,p_{12})$ in (\ref{equ_Pimixbundle}), we have the following optimization result.

\begin{pro}\label{pro_sep_mb} The optimal hybrid pricing decides $p_1^*=1-\frac{c_2}{2}$ and $p_{12}^*=1$ to include all users in the market. The equilibrium device demand and bundle demand are such that $D_1^*+D_{12}^*=1$ with $D_1^*=\frac{c_2}{2}$, $D_{12}^*=1-\frac{c_2}{2}$.
The resulting optimal profit is
\bee\label{equ_sep_mb_opR} \Pi_h(p_1^*,p_{12}^*)=1-c_1-c_2+\frac{c_2^2}{4}, \ene
which is strictly larger than the profit of bundled pricing in (\ref{equ_connectivity_pib*}).
\end{pro}

The proof is given in Appendix \ref{app_pro_sep_mb}. According to Proposition \ref{pro_sep_mb}, the optimal device price $p_1^*$ decreases with service cost $c_2$. Intuitively, as $c_2$ increases, to keep marginal profit of service selling high, the provider should stimulate more device buyers and stronger network externality, motivating a smaller device price $p_1^*$. By comparing bundled and hybrid pricing in Propositions \ref{pro_wifi_bundle} and \ref{pro_sep_mb}, the hybrid and bundled pricing have the same $p_{12}^*=1$. But for those users $D_1^*$ purchasing device-only in hybrid pricing, the marginal profit of device-selling $p_1^*-c_1=1-c_1-\frac{c_2}{2}$ is higher than the bundled pricing's uniform marginal profit $1-c_1-c_2$. It is not efficient to serve users of low mobility and incur service cost.

By further comparing the profit of hybrid pricing against that of separate pricing, we have the following result.

\begin{thm}\label{thm_connectivity_compare} Among three pricing choices, the provider prefers to adopt hybrid pricing if
\been c_1+c_2\leq 1. \enen
Otherwise, the provider chooses the separate pricing.
\end{thm}

\newcounter{myfig2}
\begin{figure}
\setcounter{myfig2}{\value{figure}}
\setcounter{figure}{3}
\centering\includegraphics[scale=0.48]{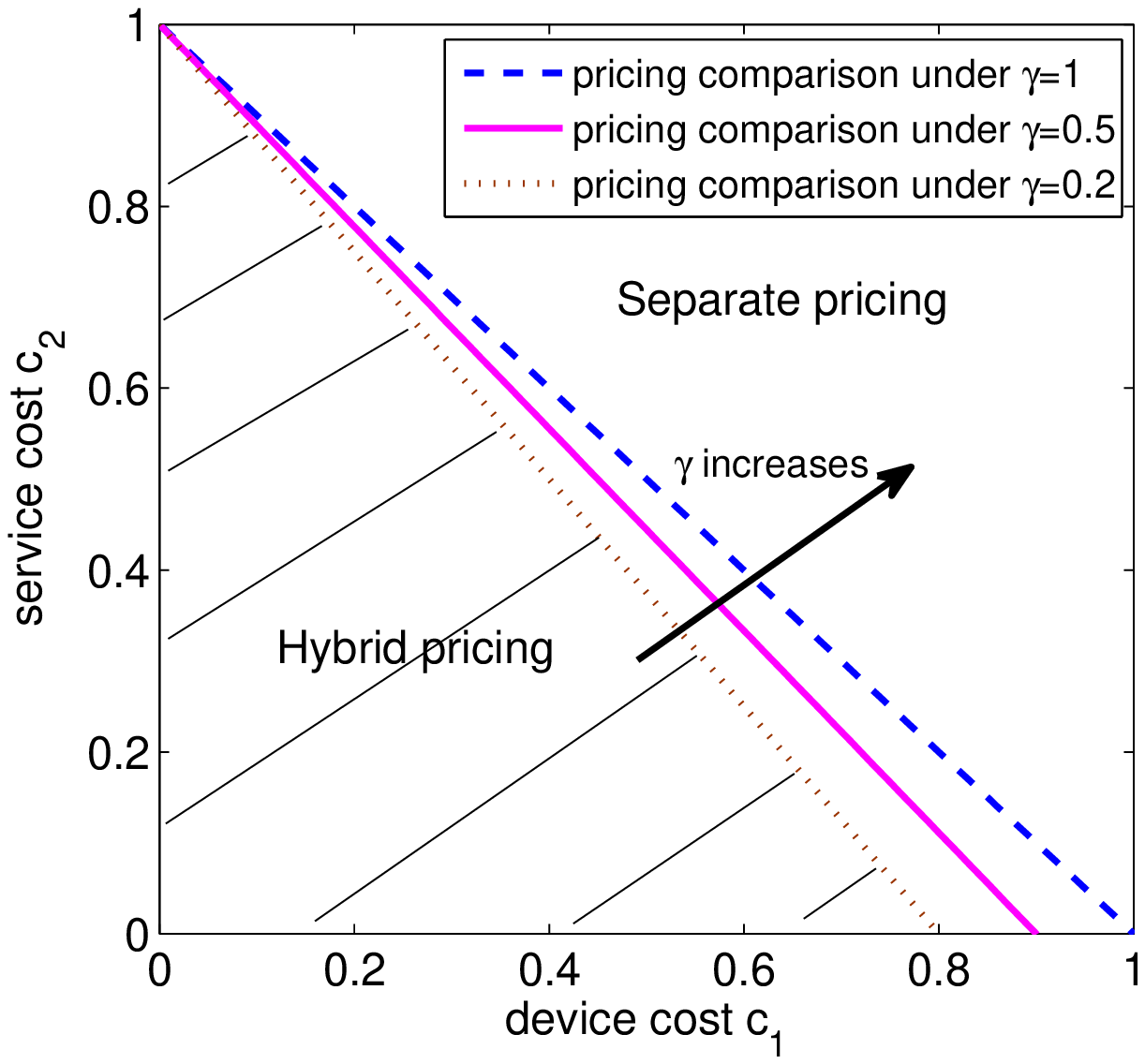}\caption{Comparison between separate pricing, bundled pricing and hybrid pricing for connectivity sharing service under different costs and service quality sensitivity $\gamma$ given in (\ref{equ_QoSgamma})}\label{compare_connectivity_mb}
\setcounter{figure}{\value{myfig2}}
\end{figure}

\newcounter{myfig3}
\setcounter{myfig3}{\value{figure}}
\setcounter{figure}{5}

With strict improvement from bundled pricing, hybrid pricing outperforms separate pricing in a larger $(c_1, c_2)$ area by comparing Figs. \ref{compare_connectivity} and \ref{compare_connectivity_mb} when $\gamma=1$. Unlike bundled pricing, the hybrid pricing still offers service when $c_2$ is high (even close to 1), as the provider can efficiently include device-only buyers to help support service value. When both device cost $c_1$ and service cost $c_2$ are high, the separate pricing outperforms the hybrid pricing as the provider can flexibly cease to provide the costly service and provides device-only option. Fig. \ref{compare_connectivity_mb} further shows that when $\gamma<1$, hybrid pricing still outperforms separate pricing in low costs area. As $\gamma$ increases (from 0.2 to 0.5 and then to 1) and the service quality $Q(D_s)$ increases more quickly with large $D_s$, hybrid pricing outperforms separate pricing even in the high device cost area. With high device cost and small $\gamma$, the provider can choose to sell less devices without damaging the service quality due to the slow increasement of network effect and service quality.%i.e., large device cost and small service cost area.
%in a larger $(c_1, c_2)$ area, i.e., additional

\begin{figure*}[t]
\centering
\subfigure[Low price regime $p_{12}\leq 1+\lambda D_{12}$]{\label{mixed1}
\begin{minipage}{.48\textwidth}
\includegraphics[width=1\textwidth]{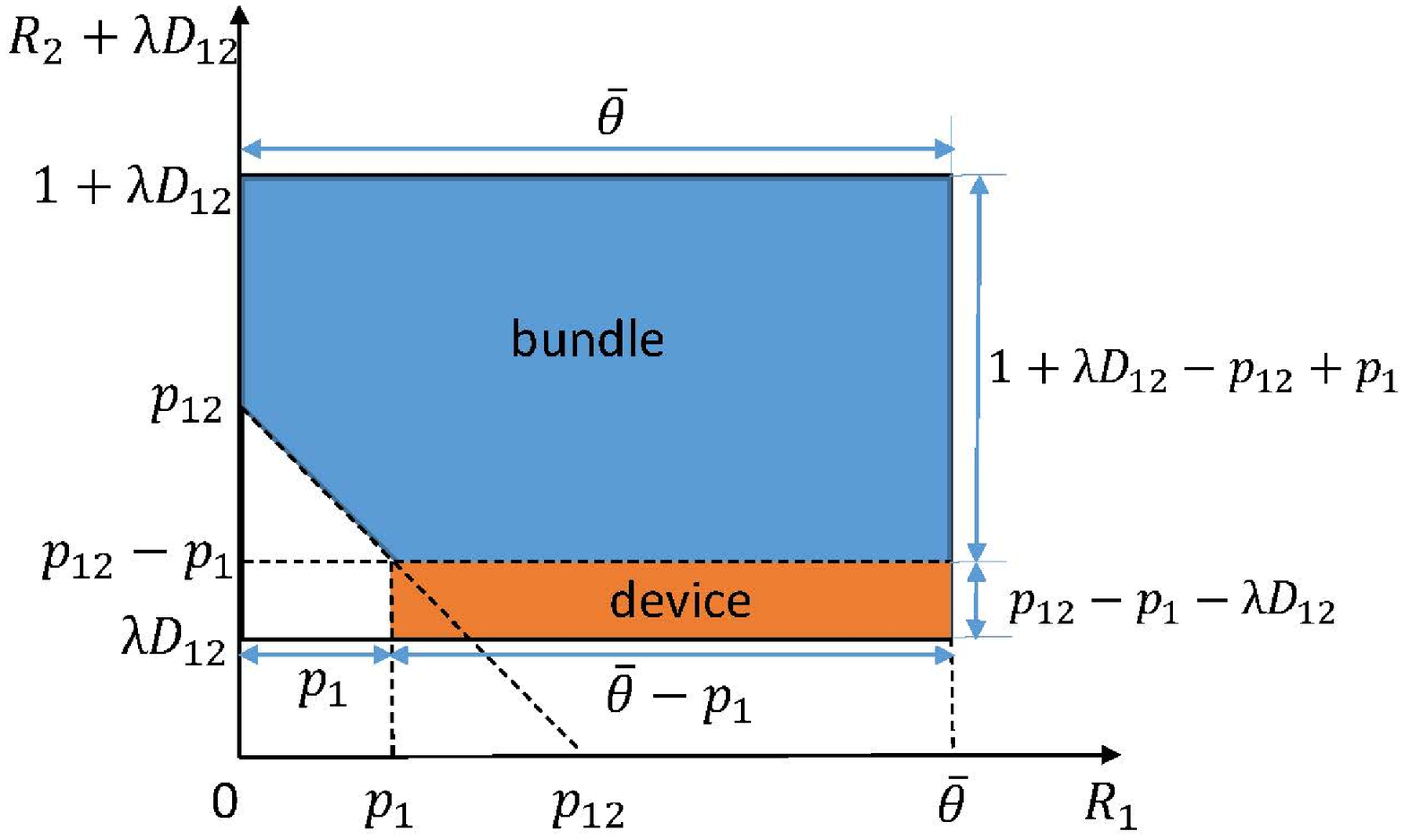}
\end{minipage}
}
\subfigure[High price regime $p_{12}>1+\lambda D_{12}$]{\label{mixed2}
\begin{minipage}{.4\textwidth}
\includegraphics[width=1\textwidth]{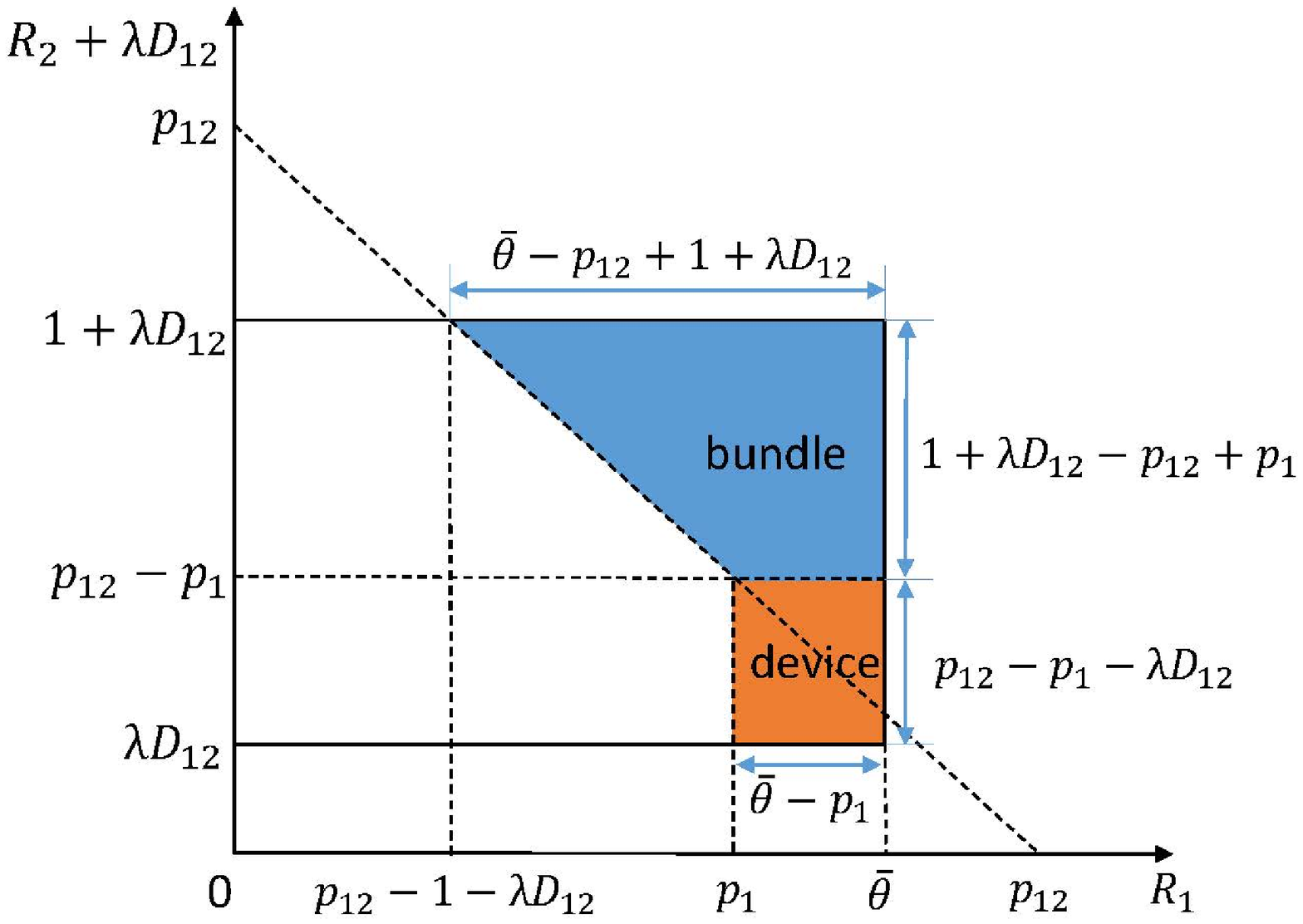}
\end{minipage}
}
\caption{Demand analysis for hybrid pricing in the content sharing service model}\label{hybridfig}
\end{figure*}

\section{Optimal Pricing for Content Sharing Services}\label{sec_contentsharing}

%\footnote{Our linear model can be extended to nonlinear functions. Compared to the linear or superlinear functions, a sublinear QoS function exhibits weaker network externality and the service attracts less subscribers.}

Different from the connectivity sharing service model, in the content sharing service model, service subscribers in a social app/service directly contribute to the network externality and users are flexible to use device and service at any time. We assume a linear service quality function $Q(D_s)=D_s$ to model one's gain from the others' virtual content sharing. As explained in Section \ref{sec_systemmodel}, a user's device and service valuations are respectively $R_1$ and $R_2+\lambda D_s$, where $\lambda$ is the degree of network externality and can be reasonably large. This is different from the bounded network externality in the physical connectivity sharing model with comparable device and service values (for indoor and outdoor usages). We assume the device value $R_1$ and service's intrinsic value $R_2$ independently follow uniform distributions in $[0,\bar{\theta}]$ and $[0,1]$, respectively. Our results can be extended to normal distribution for
both $R_1$ and $R_2$ without major change of engineering insights (see
Section \ref{sec_normal_content} for details). Practically, a user attaches a larger value to device like iWatch than the service's local benefit. In the following, we focus on the case when $\bt\geq 1$, though our analysis method also works for the other case. %\footnote{If $\bt<1$, the analysis is similar after minor modifications.}

As the content sharing services are no longer attractive to users without wearable devices, separate pricing does not work in this section. In the following, we study the provider's profit maximization under the left bundled and hybrid pricing strategies.

\subsection{Bundled Pricing for Content Sharing Services}\label{sec_content_bundle}

Under bundled pricing, the provider only offers a bundle price $p_{12}$ for both the device and service. In the content sharing service model, the users who buy the bundle product contribute to the service's QoS, i.e., $D_s=D_{12}$. By substituting $f(Q(D_{12}))=R_2+\lambda D_{12}$ into (\ref{equ_bundle}), the utility of a user with particular device and service valuations ($R_1$ and $R_2$) is
\bee\label{equ_content_bundle} U_{12}=R_1+R_2+\lambda D_{12}-p_{12}. \ene

The equilibrium demand $D_{12}^*$ depends on the bundle price $p_{12}$. If $p_{12}\geq \bt+1+\lambda D_{12}$, no user will subscribe to the bundle product. In the following, we divide the analysis for non-trivial $p_{12}<\bt+1+\lambda D_{12}$ into three cases as shown in Fig. \ref{figBundled}.

In high price regime ($\bt+\lambda D_{12}<p_{12}<\bt+1+\lambda D_{12}$ or simply $D_{12}<\frac{1}{2\bt}$), the users' subscription and their demand are shown in Fig. \ref{pure1}. A normalized unit mass of users are uniformly distributed in the whole square $[0, \bar{\theta}]\times [\lambda D_{12}, 1+\lambda D_{12}]$ with area $\bar{\theta}$. By calculating the area of the equal-right triangle, the equilibrium demand can be derived from
\bee\label{equ_d12(i)} D_{12}=\frac{(\bt-(p_{12}-1-\lambda D_{12}))^2}{2\bt}. \ene

From (\ref{equ_d12(i)}), we can show $p_{12}$ as a unique function of demand $D_{12}$, that is
\bee\label{equ_p12i} p_{12}^*(D_{12})=\bt+1+\lambda D_{12}-\sqrt{2\bt D_{12}}. \ene
Substitute it to profit $\Pi_b=(p_{12}-c_1-c_2)D_{12}$, the provider's optimization problem in Stage I is
\been \max_{D_{12}}\Pi_b(D_{12})=(\bt+1+\lambda D_{12}-\sqrt{2\bt D_{12}}-c_1-c_2)D_{12}. \enen

The objective above is a concave function of $D_{12}$, and by checking the first-order condition, we obtain the equilibrium demand as
\bee\label{equ_pro_iD} D_{12}^*=\frac{8(\bt+1-c_1-c_2)^2}{(3\sqrt{\bt}+\sqrt{9\bt-16\lambda(\bt+1-c_1-c_2)})^2}, \ene
which should still satisfy our initial assumption $D_{12}^*<\frac{1}{2\bt}$. This requires the following condition holds:
\bee 2\bt (c_1+c_2)+\bt>2(\bt^2+\lambda).\ene

The analysis for medium price regime and low price regime is similar to the high price regime case and is given in Appendix \ref{app_pro_bundle}. Then, we summarize our results in the three cases as follows.

\begin{pro}\label{pro_bundle} Under bundled pricing, the optimal price and equilibrium demand are given as follows:
\begin{itemize}
  \item If $2\bt (c_1+c_2)+\bt>2(\bt^2+\lambda)$, the provider decides high price $p_{12}\in(\bt+\lambda D_{12},\bt+1+\lambda D_{12})$ in Fig. \ref{pure1}. The equilibrium demand $D_{12}^*$ and the optimal price $p_{12}^*$ are as defined in (\ref{equ_pro_iD}) and (\ref{equ_p12i}), respectively.
  %\been p_{12}^*=\frac{\bt+1+2(c_1+c_2)-\lambda D_{12}^*}{3}. \enen

  \item If $2\bt (c_1+c_2)+\bt\leq 2(\bt^2+\lambda)$ and $2(\bt^2+\bt (c_1+c_2)+\lambda)\geq 3\bt+4\bt\lambda$, the provider decides medium price $p_{12}\in(1+\lambda D_{12},\bt+\lambda D_{12})$ in Fig. \ref{pure2}. The equilibrium demand $D_{12}^*$ is \bee\label{equ_content_iiD} D_{12}^*=\frac{\bt-c_1-c_2+\frac{1}{2}}{2(\bt-\lambda)}, \ene
      and the optimal price $p_{12}^*$ is
      \bee\label{equ_p12(ii)} p_{12}^*=\frac{2\bt+2(c_1+c_2)+1}{4}. \ene

  \item If $2(\bt^2+\bt (c_1+c_2)+\lambda)<3\bt+4\bt\lambda$, the provider decides low price $p_{12}\in(\lambda D_{12},1+\lambda D_{12})$ in Fig. \ref{pure3}. The equilibrium demand $D_{12}^*$ is the unique solution to \bee\begin{split}\label{equ_content_iiiD} &2\sqrt{2\bt}\lambda(1-D_{12})^{\frac{3}{2}}-(2\lambda-c_1-c_2)\sqrt{2\bt}\sqrt{1-D_{12}}\\
  &-3\bt(1-D_{12})+\bt=0, \end{split}\ene and the optimal price is
  \bee\label{equ_p12(iii)} p_{12}^*(D_{12})=\sqrt{2\bt(1-D_{12})}+\lambda D_{12}. \ene
\end{itemize}
\end{pro}

%\newcounter{mytempeqncnt2}
%\begin{figure*}[ht]
%\setcounter{mytempeqncnt2}{\value{equation}}
%\setcounter{equation}{50}
%\bee\begin{split}\label{equ_d12_content} &D_{12,1/2}(p_1,p_{12})=\frac{2\bt(1-\lambda)+2p_{12}\lambda-1-\lambda}{2\lambda^2}\\
%&\pm\frac{\sqrt{(2\bt(1-\lambda)+2p_{12}\lambda-1-\lambda)^2-4\lambda^2(2\bt(1-p_{12}+p_1)-2p_{12}+(p_{12}-p_1)(p_{12}+p_1)+1)}}{2\lambda^2}. \end{split}\ene
%\setcounter{equation}{\value{mytempeqncnt2}}
%\hrulefill
%\end{figure*}

\subsection{Hybrid Pricing for Content Sharing Services}\label{sec_content_hybrid}

Under hybrid pricing, the provider offers the device and bundle product simultaneously, which means that the users can buy the device only or pay additional service fee $p_{12}-p_1$ to buy both device and device-supported service. Given the device prices $p_1$ and bundle price $p_{12}$, the utility of a user for buying device only is given in (\ref{equ_Ueinitial}), and the utility for buying the bundle is given in (\ref{equ_content_bundle}).

A user will buy the device if $U_1\geq 0$. Furthermore, the user will buy the bundle if the service utility is positive, i.e., $U_{12}\geq 0$ and $R_2+\lambda D_{12}>p_{12}-p_1$. Note that even if $U_1<0$, the user will still buy the bundle once $R_2+\lambda D_{12}$ is large enough to cover the negative $U_1$.
%To be specific, users with $U_{12}\geq 0$ and $R_2+\lambda D_{12}>p_{12}-p_1$ will buy the bundle, while users with $U_1\geq 0$ and $R_2+\lambda D_{12}\leq p_{12}-p_1$ will buy device only.

Since $R_1, R_2$ follow uniform distributions, the equilibrium demand for device only is shown as the orange rectangle in Fig. \ref{mixed1} or Fig. \ref{mixed2}. By calculating its area, we have
\bee\label{equ_D1sep} D_1^*=\frac{(\bar{\theta}-p_1)(p_{12}-p_1-\lambda D_{12})}{\bt}. \ene
Note that if $p_{12}-p_1>1+\lambda D_{12}$, no one will buy the bundle product, which means that the provider sells device only at price $p_1^*=\frac{\bt+c_1}{2}$ and profit $\Pi_1=\frac{(\bt-c_1)^2}{4\bt}$.

By providing the bundle option, the equilibrium demand $D_{12}^*$ for bundle is different according to the bundle price $p_{12}$. As illustrated in Fig. \ref{mixed1} and Fig. \ref{mixed2}, we will analyze the optimal prices ($p_1^*,p_{12}^*$) under the two cases.

%\begin{figure}
%\centering\includegraphics[scale=0.38]{mixedbundledemand2}
%\textbf{Figure 6(a): Demand analysis for hybrid pricing in low price regime $p_{12}\leq 1+\lambda D_{12}$}
%\end{figure}

(i) In low price regime ($p_{12}\leq 1+\lambda D_{12}$), the users' bundle demand is shown as the blue polygon in Fig. \ref{mixed1}. We calculate the polygon area by subtracting the normalized equal-right triangle area (of side length $p_{1}$) from the rectangle $[0,\bt]\times [p_{12}-p_1,1+\lambda D_{12}]$. The equilibrium bundle demand satisfies
\bee\label{equ_2D2sep} D_{12}=\frac{\bt(1+\lambda D_{12}-(p_{12}-p_1))-\frac{p_1^2}{2}}{\bt}, \ene
which uniquely provides
\bee D_{12}^*(p_1,p_{12})=\frac{1-p_{12}+p_1-\frac{p_1^2}{2\bt}}{1-\lambda}. \ene

By substituting $D_1^*$ and $D_{12}^*$ above to $\Pi_h$ in (\ref{equ_Pimixbundle}), we can show that the profit is a concave function of $p_{12}$. The first-order condition provides the optimal $p_{12}^*(p_1)$ as a function of $p_1$, i.e.,
\bee\label{equ_sep_content_p12^*} p_{12}^*(p_1)=\frac{1+c_2}{2}+\frac{(2c_1-3p_1)p_1}{4\bt}+p_1. \ene
Then, we can rewrite $D_1^*, D_{12}^*$ and $\Pi_h$ as functions of $p_1$ only:
\bee\label{equ_pim1} \max_{p_1}\Pi_h(p_1) \ene
subject to \bee\label{equ_condition_p1} p_{12}^*(p_1)\leq 1+\lambda D_{12}^*(p_1), \ene
\bee\label{equ_condition_p1_2} p_{12}^*(p_1)-p_1-\lambda D_{12}^*(p_1)\geq 0, \ene
where the constraint (\ref{equ_condition_p1}) tells the low price regime. The left-hand-side of constraint (\ref{equ_condition_p1_2}) is the height of the orange rectangle in Fig. \ref{mixed1}, which ensures the device-only demand is non-zero, otherwise the hybrid pricing is the same as the bundled pricing.

Still, the new profit objective in (\ref{equ_pim1}) is non-convex. As only one decision variable is included, we can apply a one-dimensional exhaustive search to find optimal $p_1^*$.

%\begin{figure}
%\centering\includegraphics[scale=0.38]{mixedbundledemand}
%\textbf{Figure 6(b): Demand analysis for hybrid pricing in high price regime $p_{12}>1+\lambda D_{12}$}
%\end{figure}

(ii) In high price regime ($p_{12}>1+\lambda D_{12}$), the bundle demand is shown as the blue trapezoid area in Fig. \ref{mixed2}, which can be calculated as
\bee\begin{split}\label{equ_1D2sep} &D_{12}\\=&\frac{((\bt-p_1)+(\bt-p_{12}+1+\lambda D_{12}))(1+\lambda D_{12}-p_{12}+p_1)}{2\bt}. \end{split}\ene

Note that (\ref{equ_1D2sep}) has two roots. Denote the larger root of (\ref{equ_1D2sep}) as $D_{12,1}$ and the smaller root as $D_{12,2}$. The equilibrium demand $D_{12}$ is determined according to $p_1$ and $p_{12}$ for maximum profit:
\been
D_{12}^*(p_1,p_{12})=\begin{cases}
D_{12,1}, \text{if} ~~p_{12}-c_1-c_2>\frac{\lambda(p_1-c_1)(\bt-p_1)}{\bt};\\
D_{12,2}, \text{if} ~~p_{12}-c_1-c_2\leq\frac{\lambda(p_1-c_1)(\bt-p_1)}{\bt}.
\end{cases}
\enen

By substituting $D_1^*$ in (\ref{equ_D1sep}) and $D_{12}^*$ above into $\Pi_h$, $\Pi_h$ is just a function of $p_1$ and $p_{12}$ and the optimization problem is
%\newcounter{mytempeqncnt3}
%\setcounter{mytempeqncnt3}{\value{equation}}
%\setcounter{equation}{50}
\bee\label{equ_sepnonconvex} \max_{p_1,p_{12}}\Pi_h(p_1, p_{12}) \ene
subject to \bee\label{equ_cc1} p_{12}>1+\lambda D_{12}^*(p_1,p_{12}), \ene
\bee p_{12}-p_1-\lambda D_{12}^*(p_1,p_{12})\geq 0, \ene
where constraint (\ref{equ_cc1}) tells the high price regime. Note that (\ref{equ_sepnonconvex}) is still a non-convex problem, and we will solve it numerically below.

\subsection{Comparison between Bundled and Hybrid Pricing for Content Sharing Services}\label{sec_content_compare}

\begin{figure}
\centering\includegraphics[scale=0.34]{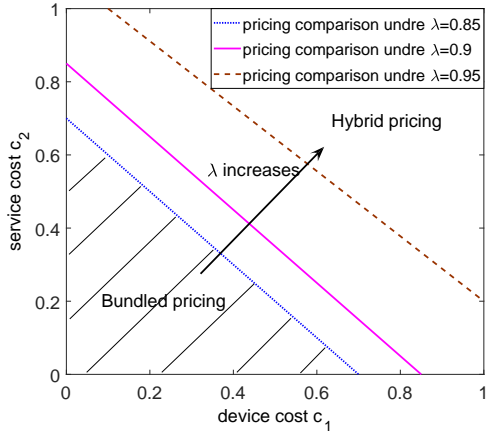}
\caption{Comparison between bundled and hybrid pricing for different $\lambda$ when $\bt=1.5$}\label{compare_c}
\end{figure}

Recall that in Section \ref{sec_connectivity_hybrid}, we show hybrid pricing strictly dominates bundled pricing. Mathematically, bundled pricing is a special case of hybrid pricing when no user buys device only. In this subsection, we analyze when the hybrid pricing degenerates to bundled pricing and numerically compare the provider's optimal profits under hybrid and bundled pricing for more general cases which cannot be solved in closed-form.%We first analyze the provider's optimal pricing strategy under strong network externality. Then, we

\newcounter{myfig8}
\begin{figure*}[t]
\setcounter{myfig8}{\value{figure}}
\setcounter{figure}{8}
\centering
\subfigure[Low price regime $\frac{p_{12}}{\omega}\leq 1+\lambda D_{12}$]{\label{omegademand1}
\begin{minipage}{.4\textwidth}
\includegraphics[width=1\textwidth]{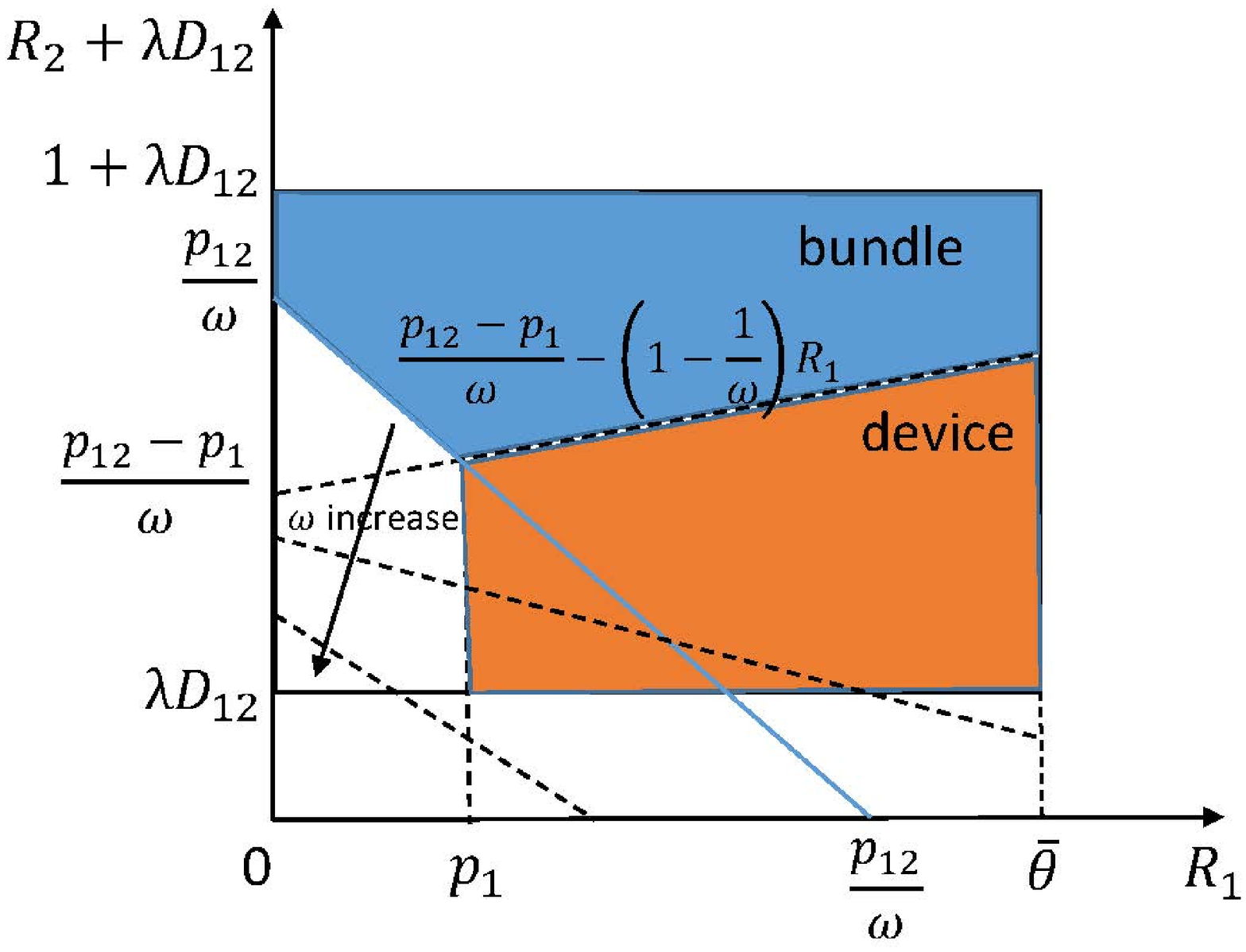}
\end{minipage}
}
\subfigure[High price regime $\frac{p_{12}}{\omega}>1+\lambda D_{12}$]{\label{omegademand2}
\begin{minipage}{.39\textwidth}
\includegraphics[width=1\textwidth]{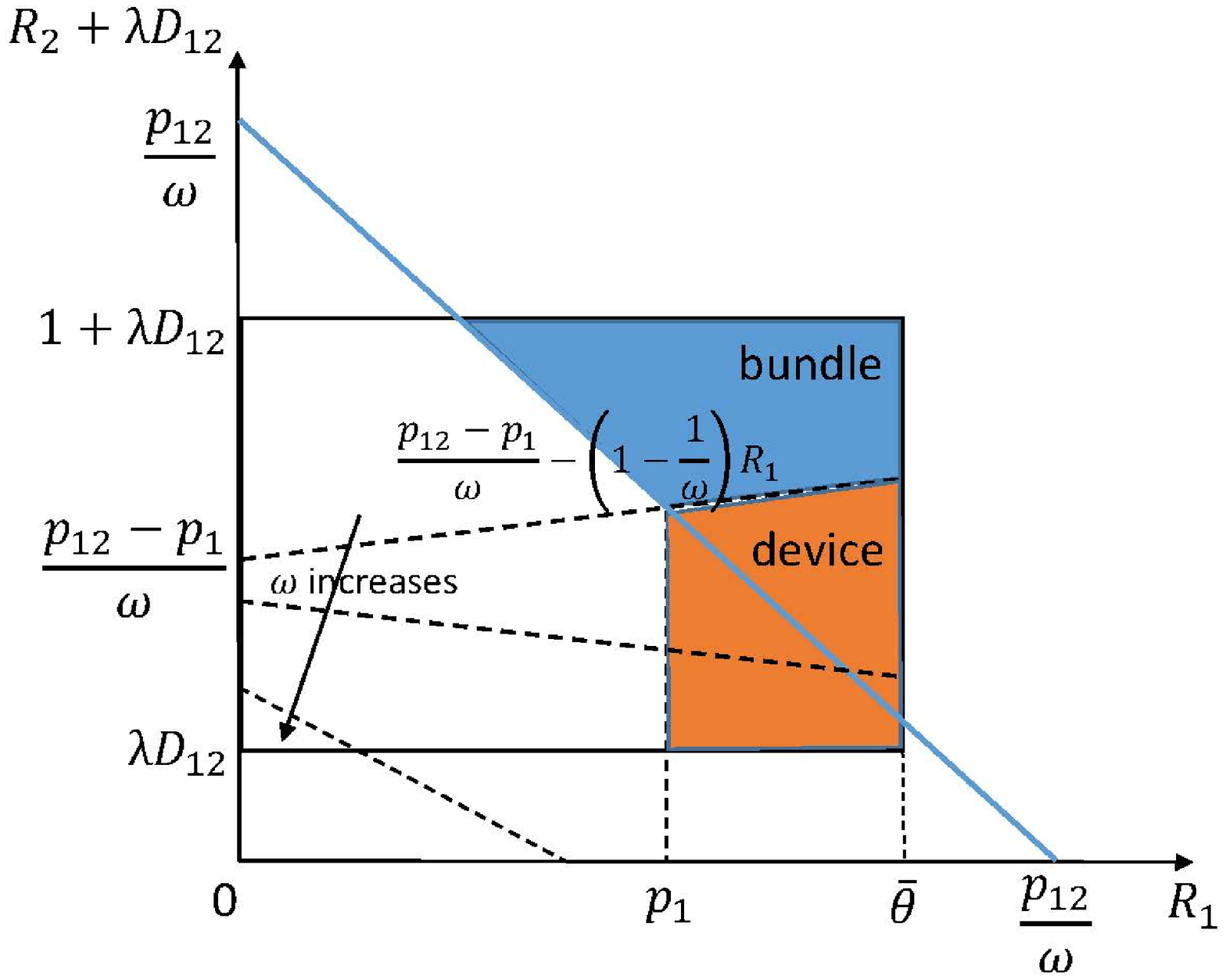}
\end{minipage}
}
\caption{Demand analysis for hybrid pricing with correlation factor $\omega$. The three dashlines in each subfigure are drawn according to the same function $\frac{p_{12}-p_1}{\omega}-(1-\frac{1}{\omega})R_1$ with different values of $\omega$}\label{hybridfigomega}
\setcounter{figure}{\value{myfig8}}
\end{figure*}

\begin{pro}\label{pro_4.1} As the degree of network externality $\lambda$ increases, the provider is more likely to sell the device and service as a bundle. To be specific, when $\lambda>\frac{3}{2}$ and $4c_1+c_2<1$, hybrid pricing degenerates to bundled pricing.
\end{pro}
%As the degree of network externality $\lambda$ increases, the provider is more likely to sell the device and service as a bundle, as illustrated in Fig. \ref{compare_c}. To be specific, if

The proof is given in Appendix \ref{app_pro_4.1}. Note that the bound is not tight due to the non-trackable optimal prices under hybrid pricing. As illustrated in Fig. \ref{compare_c}, as the network externality $\lambda$ increases (from 0.85 to 0.9 and then to 0.95), bundled pricing outperforms hybrid pricing in a larger $(c_1,c_2)$ area. As the network externality $\lambda$ increases, the service value dominates the device value, and it is better to skip device-only option and use bundled pricing to maximumly stimulate network externality, which results in a higher bundled price and thus profit. Fig. \ref{compare_c} also shows that the provider chooses bundled pricing only when the total cost is small to efficiently promote the strong network externality. If $c_1$ or $c_2$ is high, the provider prefers hybrid pricing to charge differently in device and service-related items.

\newcounter{myfig6}
\begin{figure}
\setcounter{myfig6}{\value{figure}}
\setcounter{figure}{7}
\centering\includegraphics[scale=0.32]{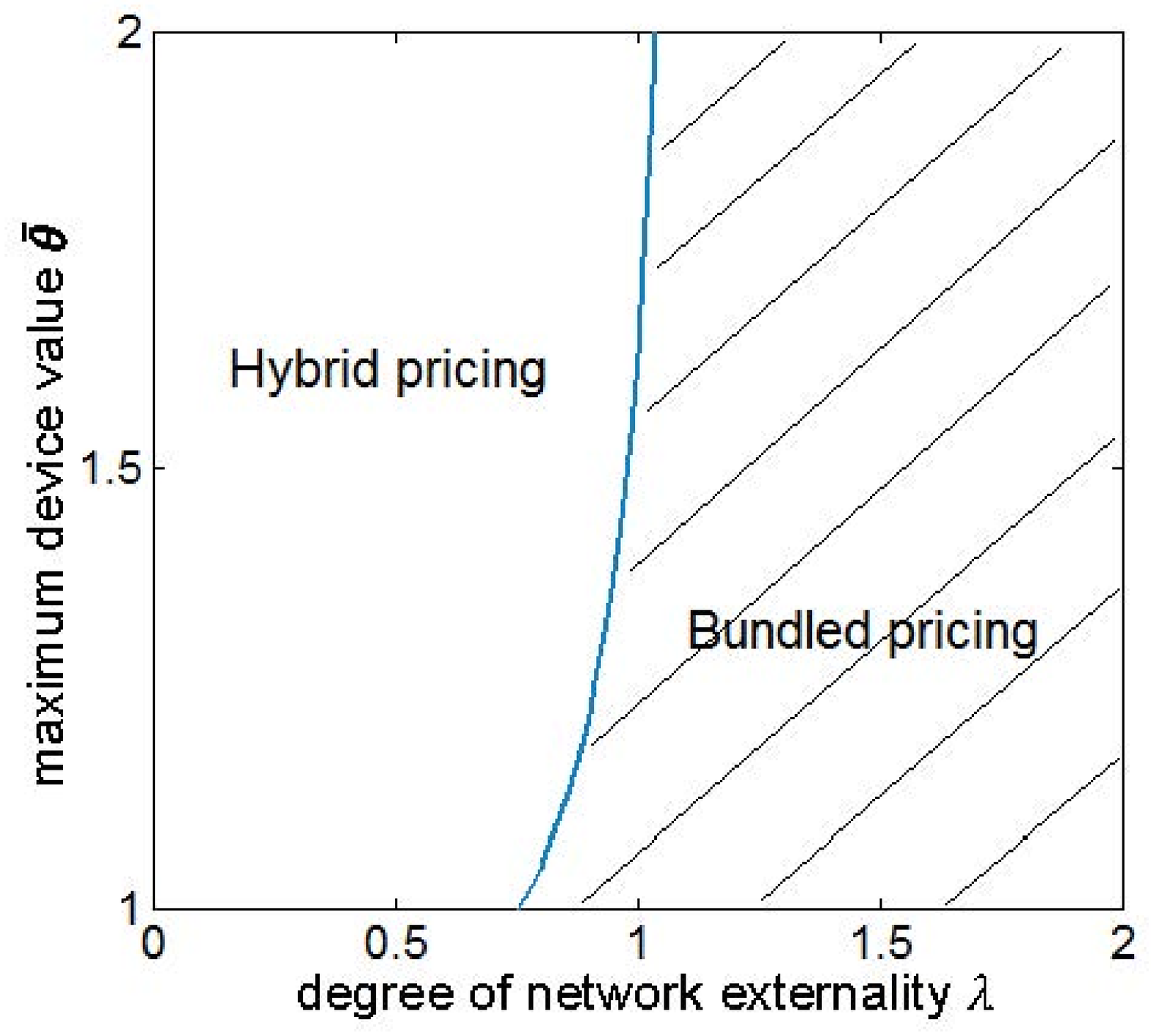}
\caption{Comparison between bundled pricing and hybrid pricing when $c_1=0.3$ and $c_2=0.1$}\label{Comparelambda}
\setcounter{figure}{\value{myfig6}}
\end{figure}

\newcounter{myfig7}
\setcounter{myfig7}{\value{figure}}
\setcounter{figure}{10}

According to Proposition \ref{pro_4.1}, when the network externality $\lambda$ is strong (larger than $\frac{3}{2}$), it is better to use bundled pricing to stimulate network externality if the total cost of device and service is low. Hence, the service valuation $\lambda D_{12}$ is high and thus the provider can charge a high bundled price. As shown in Fig. \ref{Comparelambda}, bundled pricing is equally profitable as hybrid pricing when $\lambda$ is large, i.e., hybrid pricing degenerates to bundled pricing. This is different from first physical connectivity sharing model, where device and service values for indoor and outdoor WiFi usage are related and comparable. In the virtual content sharing model, the service value is independent of device value and can be fairly large. On the other hand, when the maximum device value $\bt$ is large under small $\lambda$, the provider prefers hybrid pricing to flexibly charge a high device price to meet high device value. In this case, the bundled pricing to maximumly promote service value will inefficiently lower the marginal profit of device selling.

\section{Generalization of Our Models and Results}\label{sec_extends}

Our model can be extended to incorporate the correlation modelling of device and service valuations (subadditive or superadditive) for the content sharing service model. Other than considering the uniform distribution, we also extend to normal distributions for the valuations in both models in this section.

\subsection{Correlation between A User's Device and Service Valuations }

\newcounter{myeq1}
\begin{figure*}[ht]
\setcounter{myeq1}{\value{equation}}
\setcounter{equation}{48}
\bee\label{equ_d12_case1_w} D_{12}=\begin{cases}
\frac{1}{2\bt}((2(1+\lambda D_{12}-\frac{p_{12}}{\omega})+p_1)p_1+(\bt-p_1)(2(1+\lambda D_{12}-\frac{p_{12}}{\omega})+(1+\frac{1}{\omega})p_1+(1-\frac{1}{\omega})\bt)),\\ ~~~~~~~~~~~~~~~~~~~~~~~~~~~~~~~~~~~~~~~~~~~\text{if} ~~\frac{p_{12}}{\lambda D_{12}+1}\leq\omega<\frac{p_{12}-p_1+\bt}{\lambda D_{12}+\bt};\\
\frac{1}{2\bt}(2(1+\lambda D_{12}-\frac{p_{12}}{\omega})+p_1)p_1+\frac{\bt-p_1}{\bt}-\frac{1}{2\bt}(\frac{p_{12}}{\omega}-p_1-\lambda D_{12})(\frac{\frac{p_{12}-p_1}{\omega}-\lambda D_{12}}{1-\frac{1}{\omega}}-p_1),\\ ~~~~~~~~~~~~~~~~~~~~~~~~~~~~~~~~~~~~~~~~~~~\text{if} ~~\frac{p_{12}-p_1+\bt}{\lambda D_{12}+\bt}\leq\omega<\frac{p_{12}}{\lambda D_{12}+p_1};\\
1-\frac{(\frac{p_{12}}{\omega}-\lambda D_{12})^2}{2\bt}, ~~~~~~~~~~~~~~~~~~~~~~\text{if} ~~\omega\geq\frac{p_{12}}{\lambda D_{12}+p_1}.
\end{cases}\ene
\setcounter{equation}{\value{myeq1}}
\hrulefill
\end{figure*}

\newcounter{myeq2}
\begin{figure*}[ht]
\setcounter{myeq2}{\value{equation}}
\setcounter{equation}{49}
\bee\label{equ_d12_case2_w} D_{12}=\begin{cases}
0,~~~~~~~~~~~~~~~~~~~~~~~~~~~~~~~~~~~~~~~~~~~~~~\text{if} ~~0\leq\omega\leq\frac{p_{12}}{\lambda D_{12}+1+\bt};\\
\frac{1}{2\bt}((1+\lambda D_{12}-\frac{p_{12}}{\omega}+p_1)(p_1-\frac{p_{12}}{\omega}+1+\lambda D_{12})+(\bt-p_1)(2(1+\lambda D_{12}-\frac{p_{12}}{\omega})+(1+\frac{1}{\omega})p_1+(1-\frac{1}{\omega})\bt)),\\ ~~~~~~~~~~~~~~~~~~~~~~~~~~~~~~~~~~~~~~~~~~~~~~~~~\text{if} ~~\frac{p_{12}}{\lambda D_{12}+1+\bt}\leq\omega<\frac{p_{12}-p_1+\bt}{\lambda D_{12}+\bt};\\
\frac{1}{2\bt}(1+\lambda D_{12}-\frac{p_{12}}{\omega}+p_1)(p_1-\frac{p_{12}}{\omega}+1+\lambda D_{12})+\frac{\bt-p_1}{\bt}-\frac{1}{2\bt}(\frac{p_{12}}{\omega}-p_1-\lambda D_{12})(\frac{\frac{p_{12}-p_1}{\omega}-\lambda D_{12}}{1-\frac{1}{\omega}}-p_1),\\ ~~~~~~~~~~~~~~~~~~~~~~~~~~~~~~~~~~~~~~~~~~~~~~~~~\text{if} ~~\frac{p_{12}-p_1+\bt}{\lambda D_{12}+\bt}\leq\omega<\min(\frac{p_{12}}{\lambda D_{12}+p_1},\frac{p_{12}}{\lambda D_{12}+1});\\
\frac{(\bt-(\frac{p_{12}}{\omega}-\lambda D_{12}))+(\bt-(\frac{p_{12}}{\omega}-1-\lambda D_{12}))}{2\bt},~~~~~~~~\text{if} ~~\frac{p_{12}}{\lambda D_{12}+p_1}\leq\omega\leq\frac{p_{12}}{\lambda D_{12}+1}.
\end{cases}\ene
\setcounter{equation}{\value{myeq2}}
\hrulefill
\end{figure*}

Recall that we have assumed that users' device and service valuations are independent in Section \ref{sec_systemmodel}. In reality, users' valuations for the device and service could be subadditive or superadditive. If the device and service are substitute, the utility of purchasing both products for a user is subadditive; if the device and service are complementary, the utility of purchasing both products is superadditive. We consider the subadditivity/superadditivity correlation properties for the content sharing model in this subsection. Note that in the connectivity (WiFi) sharing model, the users are either at home or outdoor at a time, thus their device and service valuations are linearly additive (i.e., independent) according to time divisions.

Denote the correlation factor between the device and service valuations for all users as $\omega$, where $\omega>1$ tells the superadditivity property and $\omega<1$ tells the subadditivity property. In the bundled pricing, the utility of a user for buying the bundle now changes from (\ref{equ_content_bundle}) to
\bee\label{equ_content_bundle_omega} U_{12}^{\omega}=\omega(R_1+R_2+\lambda D_{12})-p_{12}. \ene

In the hybrid pricing, the utility of a user for buying device only is still given in (\ref{equ_Ueinitial}), and the utility for buying the bundle is the same as (\ref{equ_content_bundle_omega}).

Similar to the analysis in Section \ref{sec_content_bundle}, according to Fig. \ref{figBundled}, the equilibrium demand and the analysis for \textit{bundled pricing} can be divided into three cases.
\begin{itemize}
  \item If $\frac{2\bt(c_1+c_2)}{\omega}+\bt>2(\bt^2+\lambda)$, the provider decides high price $p_{12}^*(D_{12})=\omega(\bt+1+\lambda D_{12}-\sqrt{2\bt D_{12}})$, and the equilibrium demand is $D_{12}^*=\frac{8(\bt+1-\frac{c_1+c_2}{\omega})^2}{(3\sqrt{\bt}+\sqrt{9\bt-16\lambda(\bt+1-\frac{c_1+c_2}{\omega})})^2}$.

  \item If $\frac{2\bt(c_1+c_2)}{\omega}+\bt\leq 2(\bt^2+\lambda)$ and $2(\bt^2+\frac{\bt(c_1+c_2)}{\omega}+\lambda)\geq 3\bt+4\bt\lambda$, the provider decides medium price $p_{12}^*=\frac{\omega(2\bt+1)+2(c_1+c_2)}{4}$, and the equilibrium demand is $D_{12}^*=\frac{\bt+\frac{1}{2}-\frac{c_1+c_2}{\omega}}{2(\bt-\lambda)}$.

  \item If $2(\bt^2+\frac{\bt(c_1+c_2)}{\omega}+\lambda)<3\bt+4\bt\lambda$, the provider decides low price $p_{12}^*(D_{12})=w(\sqrt{2\bt(1-D_{12})}+\lambda D_{12})$, and the equilibrium demand $D_{12}^*$ is the unique solution to $2\sqrt{2\bt}\lambda(1-D_{12})^{\frac{3}{2}}-(2\lambda-\frac{c_1+c_2}{\omega})\sqrt{2\bt}\sqrt{1-D_{12}}-3\bt(1-D_{12})+\bt=0$.
\end{itemize}

Therefore, we can conclude the relationship between equilibrium demand $D_{12}^*$ and $\omega$ in the following proposition.

\begin{pro} The equilibrium bundled demand $D_{12}^*$ increases with $\omega$.
\end{pro}

For the \textit{hybrid pricing}, similar to the demand analysis illustration in Fig. \ref{hybridfig}, we also have low price regime ($\frac{p_{12}}{\omega}\leq 1+\lambda D_{12}$) and high price regime ($\frac{p_{12}}{\omega}>1+\lambda D_{12}$). Different from the equilibrium demand analysis for the independent case shown in Fig. \ref{hybridfig}, here the equilibrium demands for device only and bundled product are greatly influenced by the correlation factor $\omega$ as shown in Fig. \ref{hybridfigomega}. The division line $\frac{p_{12}-p_1}{\omega}-(1-\frac{1}{\omega})R_1$ between the bundle and device demand regions (in blue and orange) changes with $\omega$, which degenerates to the flat line in Fig. \ref{hybridfig} when $\omega=1$. As $\omega$ increases, bundled demand increases and device-only demand decreases. By calculating the orange polygon areas shown in Figs. \ref{omegademand1} and \ref{omegademand2}, the equilibrium demand for the device only changes from (\ref{equ_D1sep}) to
\bee\label{equ_D1^*omega}
D_{1}^*=\begin{cases}
\frac{\bt-p_1}{2\bt}(\frac{2p_{12}}{\omega}-(1+\frac{1}{\omega})p_1-(1-\frac{1}{\omega})\bt-2\lambda D_{12}),\\ ~~~~~~~~~~~~\text{if} ~~0\leq\omega<\frac{p_{12}-p_1+\bt}{\lambda D_{12}+\bt};\\
\frac{1}{2\bt}(\frac{p_{12}}{\omega}-p_1-\lambda D_{12})(\frac{\frac{p_{12}-p_1}{\omega}-\lambda D_{12}}{1-\frac{1}{\omega}}-p_1),\\ ~~~~~~~~~~~~\text{if} ~~\frac{p_{12}-p_1+\bt}{\lambda D_{12}+\bt}\leq\omega<\frac{p_{12}}{\lambda D_{12}+p_1};\\
0, ~~~~~~~~~\text{if} ~~\omega\geq\frac{p_{12}}{\lambda D_{12}+p_1}.
\end{cases}
\ene

%\bee
%D_{1}^*=\begin{cases}
%\frac{\bt-p_1}{2\bt}(\frac{2p_{12}}{\omega}-(1+\frac{1}{\omega})p_1-(1-\frac{1}{\omega})\bt-2\lambda D_{12}),\\ ~~~~~~~~~~~~\text{if} ~~\frac{p_{12}-p_1}{\omega}-(1-\frac{1}{\omega})\bt>\lambda D_{12};\\
%\frac{1}{2\bt}(\frac{p_{12}}{\omega}-p_1-\lambda D_{12})(\frac{\frac{p_{12}-p_1}{\omega}-\lambda D_{12}}{1-\frac{1}{\omega}}-p_1),\\ ~~~~~~~~~~~~\text{if} ~~\frac{p_{12}-p_1}{\omega}-(1-\frac{1}{\omega})\bt\leq\lambda D_{12}\\ ~~~~~~~~~~~~~~~\text{and}~~ \frac{p_{12}}{\omega}-p_1>\lambda D_{12};\\
%0, ~~~~~~~~~~\text{if} ~~\frac{p_{12}}{\omega}-p_1\leq\lambda D_{12}.
%\end{cases}
%\ene

\newcounter{myfig111}
\setcounter{myfig111}{\value{figure}}
\setcounter{figure}{9}

\begin{figure}
\centering\includegraphics[scale=0.32]{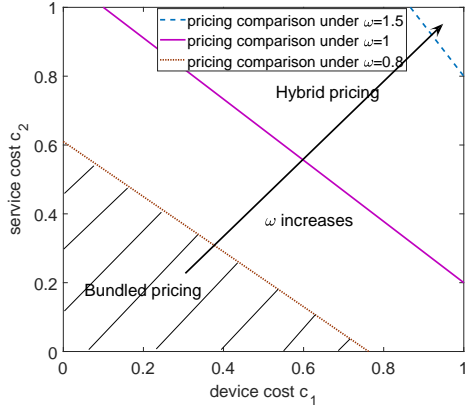}
\caption{Comparison between bundled and hybrid pricing for different $\omega$ when $\bar{\theta}=1.5$ and $\lambda=0.95$}\label{compare_w}
\end{figure}

Similar to the analysis in Section \ref{sec_content_hybrid}, the equilibrium demand analysis for the bundle product is as follows.
\begin{itemize}
  \item In low price regime ($\frac{p_{12}}{\omega}\leq 1+\lambda D_{12}$), by calculating the blue polygon area shown in Fig. \ref{omegademand1}, the equilibrium bundled demand changes from (\ref{equ_2D2sep}) to (\ref{equ_d12_case1_w}).

  \item In high price regime ($\frac{p_{12}}{\omega}>1+\lambda D_{12}$), by calculating the blue polygon area shown in Fig. \ref{omegademand2}, the equilibrium bundled demand changes from (\ref{equ_1D2sep}) to (\ref{equ_d12_case2_w}).

\end{itemize}

According to the equilibrium demands given in (\ref{equ_D1^*omega}), (\ref{equ_d12_case1_w}) and (\ref{equ_d12_case2_w}), given any device price $p_1$ and bundle price $p_{12}$, we can rigorously prove the following proposition.

\begin{pro}In the hybrid pricing, the equilibrium demand for device only $D_{1}^*(p_1,p_{12})$ decreases to $0$ as the correlation factor $\omega$ increases, while the equilibrium demand for bundle product $D_{12}^*(p_1,p_{12})$ increases with $\omega$.
\end{pro}

\newcounter{mytempeqncnt1}
\setcounter{mytempeqncnt1}{\value{equation}}
\setcounter{equation}{50}

The equilibrium demands are not in closed-form and the objective $\max_{p_1,p_{12}}\Pi_h(p_1,p_{12})$ given in (\ref{equ_Pimixbundle}) is a non-convex problem. It is difficult to theoretically analyze the optimal hybrid pricing, and we use numerical methods by exhaustively searching the optimal device and bundled prices as in Section \ref{sec_content_compare}. Then, we compare the provider's optimal profits under bundled and hybrid pricing. As illustrated in Fig. \ref{compare_w}, hybrid pricing outperforms bundled pricing in high ($c_1, c_2$) area, which is similar to Fig. \ref{compare_c}. However, as the correlation factor $\omega$ increases, hybrid pricing is more likely to degenerate to bundled pricing. Intuitively, as $\omega$ increases, users prefer to buy the device and service as a bundle to increase their utilities jointly, thus it is better for the provider to skip device-only option and use bundled pricing to maximumly stimulate network externality.

\subsection{Normal Distribution for Device and Service Valuations}

We extend our models to normal distributions for the device and service valuations in both connectivity and content sharing service models in this subsection.

% and \ref{sec_contentsharing}
\subsubsection{Connectivity Sharing Service Model}\label{sec_normal_connectivity}

Now a user's mobility factor $\alpha\in[0,1]$ follows a normal distribution with mean $\mu_{\alpha}\in(0,1)$ and variance $\sigma_{\alpha}>0$ instead of uniform distribution assumed in Section \ref{sec_trafficsharing}. This is a truncated normal distribution version, as the user mobility $\alpha$ is bounded in $[0,1]$ in practice. Similar to the analysis in Section \ref{sec_trafficsharing}, the equilibrium demand analysis for the three pricing schemes is as follows:% and $\alpha$ lies within the interval
\begin{itemize}
  \item Separate pricing: The equilibrium demand for the device changes from (\ref{equ_sep_D1*}) to
\bee\label{equ_normal_sep1} D_1^*(p_1)=\frac{1}{\Sigma}(\text{erf}(\frac{1-p_1-\mu_{\alpha}}{\sqrt{2}\sigma_{\alpha}})+\text{erf}(\frac{\mu_{\alpha}}{\sqrt{2}\sigma_{\alpha}})), \ene
and the equilibrium demand for the service changes from (\ref{equ_sep_D2*}) to
\bee\label{equ_normal_sep2}  D_2^*(p_1,p_2)=\frac{1}{\Sigma}(\text{erf}(\frac{1-\mu_{\alpha}}{\sqrt{2}\sigma_{\alpha}})-\text{erf}(\frac{\frac{p_2}{D_1^*}-\mu_{\alpha}}{\sqrt{2}\sigma_{\alpha}})), \ene
where $\Sigma$ is denoted as $\text{erf}(\frac{1-\mu_{\alpha}}{\sqrt{2}\sigma_{\alpha}})+\text{erf}(\frac{\mu_{\alpha}}{\sqrt{2}\sigma_{\alpha}})$.
  \item Bundle pricing: If (\ref{equ_normal_bundle_connectivity}) below has a real root in the region $[0,1]$, the equilibrium demand $D_{12}^*\in[0,1]$ is the unique solution to
  \bee\label{equ_normal_bundle_connectivity} D_{12}=\frac{1}{\Sigma}(\text{erf}(\frac{\frac{1-p_{12}}{1-D_{12}}-\mu_{\alpha}}{\sqrt{2}\sigma_{\alpha}})+\text{erf}(\frac{\mu_{\alpha}}{\sqrt{2}\sigma_{\alpha}})). \ene

 Otherwise, $D_{12}^*=1$.

  \item Hybrid pricing: The equilibrium demand for the device changes from (\ref{equ_sep_mb_D1}) to
\bee\label{equ_normal_hybridD1} D_1^*(p_1,p_{12})=\frac{1}{\Sigma}(\text{erf}(\frac{p_{12}-p_1-\mu_{\alpha}}{\sqrt{2}\sigma_{\alpha}})+\text{erf}(\frac{\mu_{\alpha}}{\sqrt{2}\sigma_{\alpha}})), \ene
and the equilibrium demand for the bundle changes from (\ref{equ_sep_mb_D3}) to
\bee\label{equ_normal_hybridD12} D_{12}^*(p_1,p_{12})=\frac{1}{\Sigma}(\text{erf}(\frac{1-\mu_{\alpha}}{\sqrt{2}\sigma_{\alpha}})-\text{erf}(\frac{p_{12}-p_1-\mu_{\alpha}}{\sqrt{2}\sigma_{\alpha}})). \ene

\end{itemize}

Actually, for any continuous distribution of mobility factor $\alpha$, below we can obtain similar results as Proposition \ref{pro_sep_mb} and Theorem \ref{thm_connectivity_compare} for uniform distribution.

%Note that hybrid pricing degenerates to bundled pricing if the optimal device price $p_1^*=p_{12}$, i.e., $\frac{\partial \Pi_h}{\partial p_1}|_{p_1=p_{12}}=0$ or $\frac{\partial \Pi_h}{\partial p_1}|_{p_1=p_{12}}>0$. Actually, for any distribution $f_{\alpha}$ of $\alpha$, we have $\frac{\partial \Pi_h}{\partial p_1}|_{p_1=p_{12}}=-c_2f_{\alpha}(0)<0$. Therefore, we have the following proposition, which is similar to the result in Proposition \ref{pro_sep_mb} for uniform distribution.

%$\frac{\partial \Pi_h}{\partial p_1}|_{p_1=p_{12}}=-c_2<0$. Therefore, we have the following proposition, which is similar to the result in Proposition \ref{pro_sep_mb} for uniform distribution.

\begin{pro}\label{pro_hybrid_connectivity} Hybrid pricing strictly dominates bundled pricing for any continuous distribution of $\alpha$ (including the normal distribution).
\end{pro}

The proof of Proposition \ref{pro_hybrid_connectivity} is given in Appendix \ref{app_pro_hybrid_connectivity}.

%\begin{figure}
%\centering\includegraphics[scale=0.32]{compare_connectivity_normal}
%\caption{Comparison between separate pricing, bundled pricing and hybrid pricing for connectivity sharing service when $\alpha$ follows normal distribution}\label{normalalpha}
%\end{figure}

As hybrid pricing always outperforms bundled pricing, we only need to compare separate pricing and hybrid pricing. Note that if $c_1+c_2>1$, according to (\ref{equ_Ubwifi}), the users will not buy the bundle product since their utility for buying the bundle $U_{12}(\alpha)<0$ due to $p_{12}\geq c_1+c_2>1$. Thus, the provider will always choose separate pricing to maximize its profit for selling device only. Then, we have the following proposition.

\begin{pro} For any continuous distribution of $\alpha$, the provider prefers to adopt separate pricing if
\bee c_1+c_2>1. \ene
%Otherwise, the provider choose separate pricing.
\end{pro}

%\textbf{Proof:} Denote the optimal device and service prices for separate pricing as $p_1^s$ and $p_2^s$, respectively. If $c_1+c_2>1$, according to (\ref{equ_Ubwifi}), we can see that the users will not buy the bundle product since their utility for buying the bundle $U_{12}(\alpha)<0$ due to $p_{12}\geq c_1+c_2>1$. Thus, the provider will always choose separate pricing to maximize its profit for selling device only.

%If $c_1+c_2\leq 1$, according to (\ref{equ_normal_hybridD1}) and (\ref{equ_normal_hybridD12}), we can see that all users contribute to the WiFi coverage, i.e., $D_1^*+D_{12}^*=1$. Thus, the provider would set the bundled pricing as the maximum possible value $p_{12}^*=1$ to extract all the surplus value of the users. Then, as shown in (\ref{equ_normal_sep1}) and (\ref{equ_normal_hybridD1}), the equilibrium demands for the device only in separate pricing scheme and bundled pricing scheme become identical. Therefore, the hybrid pricing is better than separate pricing as it maximizes the positive network effect to involve all the users and charges the highest bundle price $p_{12}^*=1$. \qed

%If $p_1^s+p_2^s\leq 1$, hybrid pricing is obviously better than separate pricing as the provider can charge a service price $p_{12}-p_1^s=1-p_1^s$ higher than $p_2^s$ in the separate pricing and all users are involved in the market. If $p_1^s+p_2^s>1$, users with $\alpha\leq 1-p_1<p_2$ will buy the device and users with $\alpha\geq \frac{p_2}{D_1}\geq p_2$ will buy the service.
%no user will buy the service as $U_2(\alpha)<0$ for all users due to

This result is similar to Fig. \ref{compare_connectivity_mb} for uniform distribution. When both device cost $c_1$ and service cost $c_2$ are high, the provider would choose separate pricing to flexibly cease to provide costly service and only provides device option. However, when both costs are small, hybrid pricing dominates separate pricing as it stimulates the positive network effect.

%Recall Fig. \ref{compare_connectivity_mb} for uniform distribution, when both device cost $c_1$ and service cost $c_2$ are high, the provider would choose separate pricing to flexibly cease to provide costly service and provides device-only option.

\subsubsection{Content Sharing Services}\label{sec_normal_content}

Now we turn to the content sharing model, where the users' device valuation $R_1\in[0,\bt]$ and service valuation $R_2\in[0,1]$ follow normal distributions with mean $\mu_1\in(0,\bt), \mu_2\in(0,1)$ and variance $\sigma_1,\sigma_2>0$, respectively. Here, we further investigate the impact of nonlinearity of service quality function $Q(D_{12})=D_{12}^{\gamma}$ on the provider's pricing design, where $\gamma\leq 1$ and small $\gamma$ tells that the QoS increases very quickly at the beginning and then very slowly. Similar to the analysis in Section \ref{sec_content_bundle}, according to Fig. \ref{figBundled}, the equilibrium demand analysis for \textit{bundled pricing} is divided into three cases.
\begin{itemize}
  \item In high price regime ($\bt+\lambda D_{12}^{\gamma}<p_{12}<\bt+1+\lambda D_{12}^{\gamma}$), the equilibrium bundled demand (not in closed-form) is
      \bee\label{equ_normal_bundled1}\begin{split} D_{12}=&\frac{1}{\Phi}\int_{p_{12}-\bt-\lambda D_{12}^{\gamma}}^1f_{R_2}(R_2)\psi(R_2) d R_2, \end{split}\ene

  where $f_{R_2}(R_2)=\frac{1}{\sqrt{2\pi}\sigma_2}\exp(-\frac{(R_2-\mu_2)^2}{2\sigma_2^2})$, $\psi(R_2)=\text{erf}(\frac{\bt-\mu_1}{\sqrt{2}\sigma_1})-\text{erf}(\frac{p_{12}-R_2-\lambda D_{12}^{\gamma}-\mu_1}{\sqrt{2}\sigma_1})$,
 $\Phi=\frac{1}{2}\Big(\text{erf}(\frac{\bt-\mu_1}{\sqrt{2}\sigma_1})+\text{erf}(\frac{\mu_1}{\sqrt{2}\sigma_1})\Big)\Big(\text{erf}(\frac{1-\mu_2}{\sqrt{2}\sigma_2})+\text{erf}(\frac{\mu_2}{\sqrt{2}\sigma_2})\Big)$. It is difficult to solve the integration and simplify (\ref{equ_normal_bundled1}).

  \item In medium price regime ($1+\lambda D_{12}^{\gamma}<p_{12}<\bt+\lambda D_{12}^{\gamma}$), the equilibrium bundled demand is
      \bee \begin{split} D_{12}=&\frac{1}{\Phi}\int_{0}^1f_{R_2}(R_2)\psi(R_2)d R_2. \end{split}\ene

  \item In low price regime ($\lambda D_{12}^{\gamma}<p_{12}<1+\lambda D_{12}^{\gamma}$), the equilibrium bundled demand is
      \bee \begin{split} D_{12}=&\frac{1}{\Phi}\int_{0}^{p_{12}-\lambda D_{12}^{\gamma}}f_{R_2}(R_2)\psi(R_2)d R_2.\end{split}\ene
\end{itemize}

For the \textit{hybrid pricing}, the equilibrium demand for device only is
\bee \begin{split} D_1=&\frac{1}{2\Phi}\Big(\text{erf}(\frac{\bt-\mu_1}{\sqrt{2}\sigma_1})-\text{erf}(\frac{p_1-\mu_1}{\sqrt{2}\sigma_1})\Big)\times\\ &\Big(\text{erf}(\frac{p_{12}-p_1-\lambda D_{12}^{\gamma}-\mu_2}{\sqrt{2}\sigma_2})+\text{erf}(\frac{\mu_2}{\sqrt{2}\sigma_2})\Big). \end{split}\ene
And the equilibrium demand analysis for the bundle product is divided into two cases.
\begin{itemize}
  \item In low price regime ($p_{12}\leq 1+\lambda D_{12}^{\gamma}$), the equilibrium bundled demand is
      \bee \begin{split} D_{12}=&\frac{1}{\Phi}\int_{p_{12}-p_1-\lambda D_{12}^{\gamma}}^{p_{12}-\lambda D_{12}^{\gamma}}f_{R_2}(R_2)\psi(R_2)d R_2. \end{split}\ene
  \item In high price regime ($p_{12}>1+\lambda D_{12}^{\gamma}$), the equilibrium bundled demand is
      \bee \begin{split} D_{12}=&\frac{1}{\Phi}\int_{p_{12}-p_1-\lambda D_{12}^{\gamma}}^{1}f_{R_2}(R_2)\psi(R_2) d R_2. \end{split}\ene
\end{itemize}

\begin{figure}
\centering\includegraphics[scale=0.33]{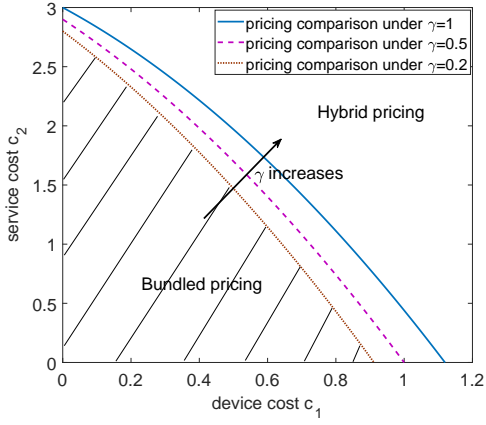}\caption{Comparison between bundled pricing and hybrid pricing for content sharing service under different service quality sensitivity $\gamma$ when device valuation $R_1\in[0,\bt]$ and service valuation $R_2\in[0,1]$ follow normal distribution}\label{compare_content_normal_diff_gamma}
\end{figure}

Substitute the equilibrium demands into the provider's total profit under bundled pricing $\Pi_b(p_{12})$ in (\ref{equ_pi_b}) and hybrid pricing $\Pi_h(p_1,p_{12})$ in (\ref{equ_Pimixbundle}), respectively, we can show that the objectives are non-convex. Therefore, we numerically search for the optimal prices and compare the profits under both pricing schemes as in Section \ref{sec_contentsharing}. Fig. \ref{compare_content_normal_diff_gamma} shows that the provider would choose hybrid pricing for large costs regime to charge differently in device and service-related items, and he would choose bundled pricing for small costs area to efficiently promote the strong network externality. This is similar to Fig. \ref{compare_c}. Fig. \ref{compare_content_normal_diff_gamma} also show the impact of $\gamma$, where a smaller $\gamma$ value exhibits weaker network externality and thus the provider is less likely to choose bundled pricing.

%Compared to the linear or superlinear functions, a sublinear QoS function exhibits weaker network externality and the service attracts less subscribers.

\section{Conclusions}

This paper has studied profit maximization for the wireless device provider in two popular service models, namely physical connectivity sharing and virtual content sharing services, under three pricing strategies -- separate pricing, bundled pricing and hybrid pricing, and gives guidance to which pricing strategy to apply. In the physical service model where the device-owners contribute to the connectivity sharing, it is shown that bundled pricing outperforms separate pricing as long as the total device and service cost is reasonably low to stimulate network externality, and hybrid pricing always dominates bundled pricing thanks to the flexibility to keep high marginal profit of device-selling. In the virtual sharing service model where only service subscribers contribute to the virtual content sharing and the network externality can be fairly strong, we show that hybrid pricing degenerates to bundled pricing when the network externality is larger than the average device valuation.

% or


\appendices
\section{Proof of Proposition \ref{pro_sep_wifi}}\label{app_pro_sep_wifi}

To find the optimal prices, we substitute the equilibrium device demand $D_1^*(p_1)$ in (\ref{equ_sep_D1*}) and service demand $D_2^*(p_1,p_2)$ in (\ref{equ_sep_D2*}) to the provider's profit $\Pi_s$ in (\ref{equ_pro}) and take derivative with respect to $p_1, p_2$, respectively. Then, we get the relationship between optimal $p_1$ and $p_2$ as
\bee\label{equ_sep_partial_p1} \frac{\partial \Pi_s}{\partial p_1}=1-2p_1+c_1-\frac{p_2(p_2-c_2)}{(1-p_1)^2}=0, \ene
\bee \frac{\partial \Pi_s}{\partial p_2}=1-\frac{p_2}{1-p_1}-\frac{p_2-c_2}{1-p_1}=0. \ene

Since $\frac{\partial^2 \Pi_s}{\partial p_2^2}<0$, the optimal profit $\Pi_s$ is concave with $p_2$ and thus the optimal service price is $p_2^*=\frac{1+c_2-p_1^*}{2}$ in (\ref{equ_sep_p2*}). If social service is provided, i.e., $D_2^*(p_1,p_2)\geq 0$, or simply $p_1+p_2\leq1$, we require $p_1\leq1-c_2$ by substituting (\ref{equ_sep_p2*}) into $p_1+p_2^*\leq 1$.

Substitute $p_2^*$ to (\ref{equ_sep_partial_p1}), we have the optimal device price as the solution to (\ref{equ_sep_solvep1}).

%\begin{figure}
%\centering\includegraphics[scale=0.5]{proof31}
%\caption{Illustration of unique real root in the reasonable device price region $[0,1-c_2]$ for $c_1+2c_2<1$}\label{proof3.1}
%\end{figure}
%
%\begin{figure}
%\centering\includegraphics[scale=0.5]{proof312roots}
%\caption{Illustration of two real roots in the reasonable device price region $[0,1-c_2]$ for $c_1+2c_2\geq1$}\label{proof3.12roots}
%\end{figure}
%
%\begin{figure}
%\centering\includegraphics[scale=0.5]{proofnoroot}
%\caption{Illustration of no root in the reasonable device price region $[0,1-c_2]$ for $c_1+2c_2\geq1$}\label{proof31noroot}
%\end{figure}

%to include the left-hand-side of (\ref{equ_sep_solvep1})
Denote the profit's first-derivative function as $f(p_1)=\frac{3}{4}-2p_1+c_1+\frac{c_2^2}{4(1-p_1)^2}$. We can see that $f(p_1)=0$ has at most three roots and $f(p_1)$ satisfies $f(p_1=0)>0$, $f(p_1=1-c_2)=c_1+2c_2-1$, $f(p_1=1)=+\infty$ and $f(p_1=+\infty)=-\infty$. In the following, we discuss the provider's optimal price design depending on the two costs.

\begin{figure*}[t]
\centering
\subfigure[Unique real root when $c_1+2c_2<1$]{\label{proof3.1}
\begin{minipage}{.3\textwidth}
\includegraphics[width=1\textwidth]{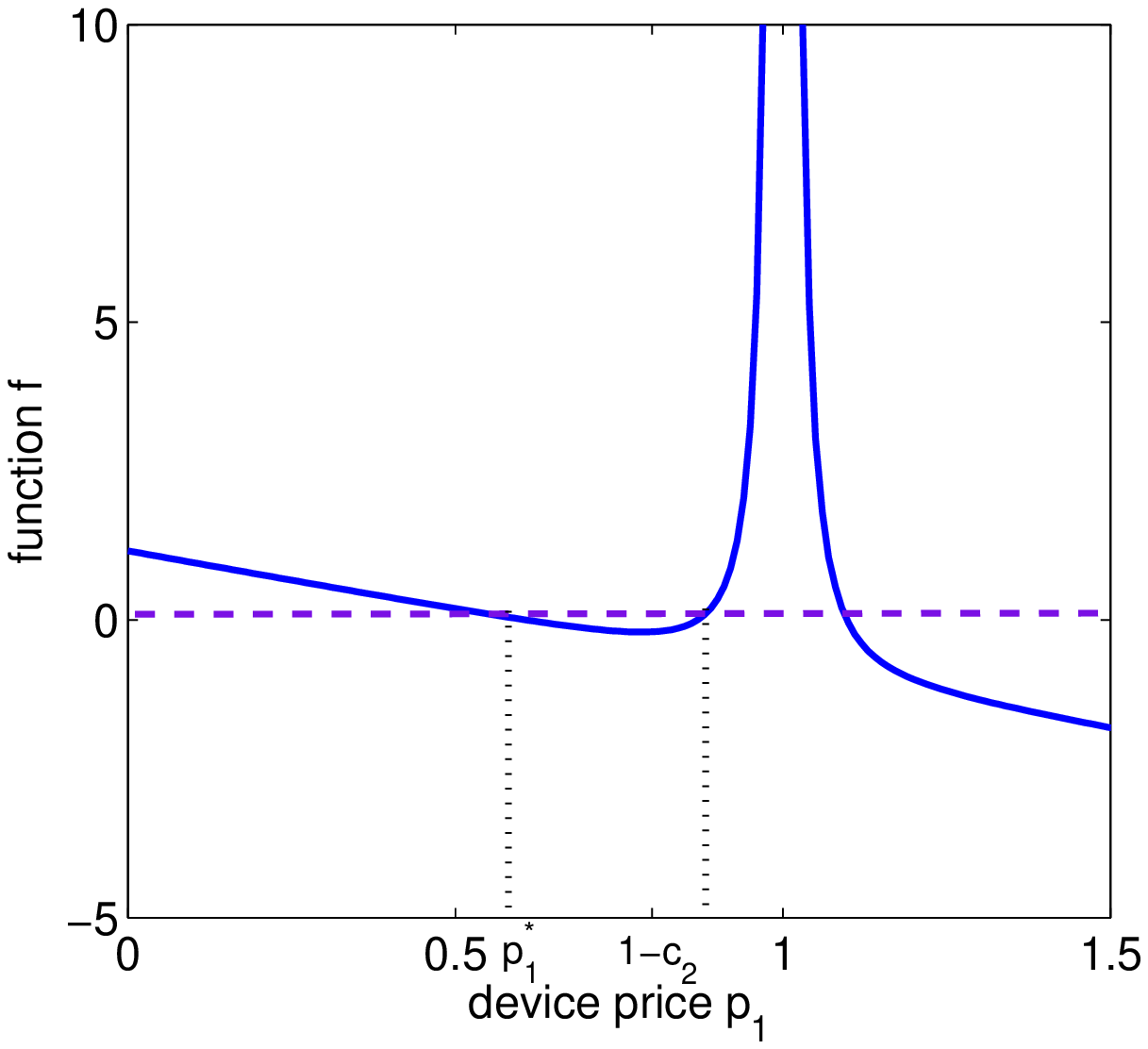}
\end{minipage}
}
\subfigure[Two real roots when $c_1+2c_2\geq1$]{\label{proof3.12roots}
\begin{minipage}{.305\textwidth}
\includegraphics[width=1\textwidth]{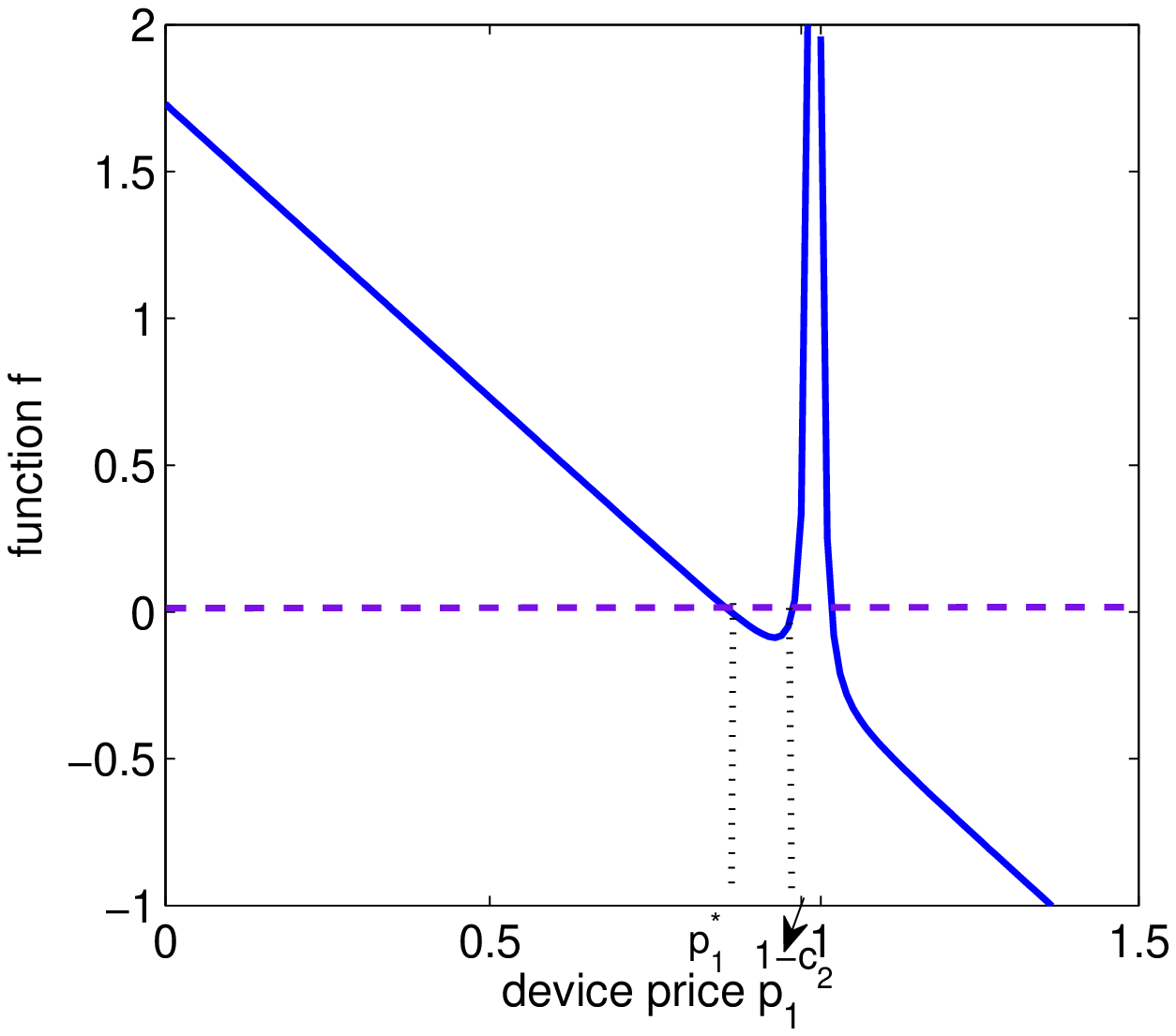}
\end{minipage}
}
\subfigure[No real root when $c_1+2c_2\geq1$]{\label{proof31noroot}
\begin{minipage}{.3\textwidth}
\includegraphics[width=1\textwidth]{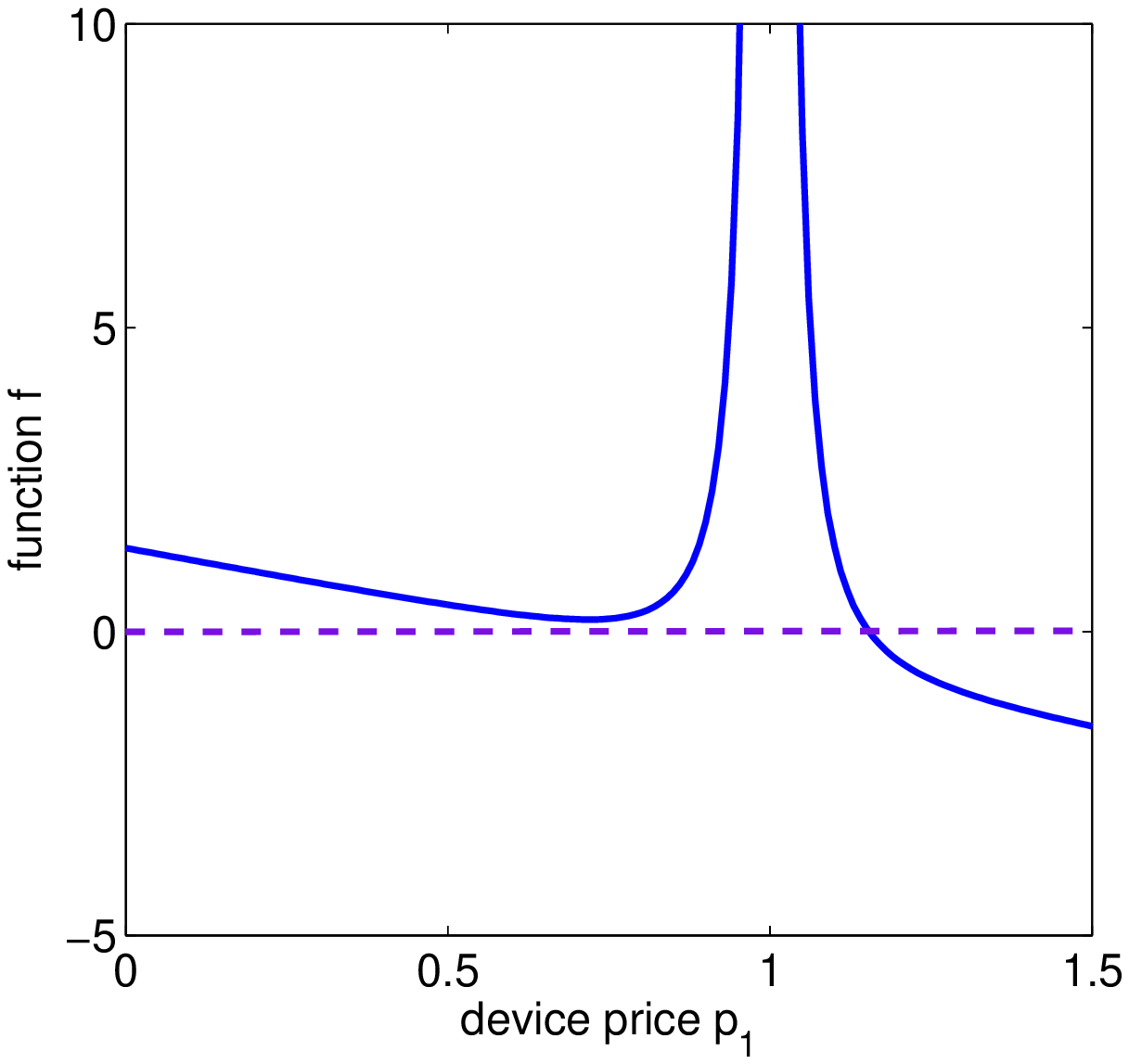}
\end{minipage}
}
\caption{Illustration of existence of real root in the reasonable device price region $[0,1-c_2]$}
\end{figure*}

%the discriminant of the cubic equation
When $f(1-c_2)=c_1+2c_2-1<0$, as shown in Fig. \ref{proof3.1}, $f(p_1)=0$ has three real roots, two of which are less than $1$. Since only the smallest root of (\ref{equ_sep_solvep1}) is in the reasonable region $[0,1-c_2]$ (see Fig. \ref{proof3.1}), we have $p_1^*$ as the minimal solution of (\ref{equ_sep_solvep1}).

When $f(1-c_2)=c_1+2c_2-1\geq0$, $f(p_1)$ has two real roots if $\mu\leq0$ or no root if $\mu>0$ in the reasonable region $p_1\in[0,1-c_2]$, where
\been \mu=\kappa^2+\Big(\frac{14+8c_1}{24}-\frac{(19+4c_1)^2}{576}\Big)^3, \enen
with \been\begin{split} \kappa=&-\frac{(19+4c_1)(14+8c_1)}{384}+\Big(\frac{19+4c_1}{24}\Big)^3\\&+\frac{c_2^2+4c_1+3}{16}. \end{split}\enen
If $f(p_1)$ has two roots as shown in Fig. \ref{proof3.12roots}, the smaller one satisfies $\frac{\partial^2 \Pi_s}{\partial p_1^2}<0$ whereas the larger one satisfies $\frac{\partial^2 \Pi_s}{\partial p_1^2}>0$. Therefore, the provider will choose the smaller one as the device price.% which is given as (\ref{equ_sep_connectivity_p1*}).

If $f(p_1)$ has no solution in the reasonable region $p_1\in[0,1-c_2]$ as shown in Fig. \ref{proof31noroot}, $f(p_1)>0$ always holds and thus $\Pi_s$ increases with $p_1$ in the region $p_1\in[0,1-c_2]$. When $p_1=1-c_2$, $p_2=c_2$ according to (\ref{equ_sep_p2*}) and the provider can't gain positive profit from the service. Hence, the provider would not provide the social service in the first place and aims to sell the device only. The profit received by selling the device only is
\bee \Pi_1(p_1)=(p_1-c_1)D_1, \ene where $D_1$ is given in (\ref{equ_sep_D1*}).
Using the first order condition, the provider sets $p_1^*=\frac{1+c_1}{2}$ and the resulting profit is $\Pi_1(p_1^*)=\frac{(1-c_1)^2}{4}$. By comparing the optimal profits with and without service, the provider chooses the pricing strategy that optimize its profit. That is, if $\Pi_s(p_1^*,p_2^*)>\Pi_1((p_1^*)$, where $\Pi_s(p_1^*,p_2^*)$ can be obtained by insert $p_1^*,p_2^*$ into (\ref{equ_pro}), the provider provides both device and service; Otherwise, the provider sells device only.
%Note that when $c_1+2c_2\geq 1$, $p_1^*=\frac{1+c_1}{2}>1-c_2$. By comparing the optimal profits with and without service, the provider chooses the smaller solution as the optimal device price or not to provide the service when $c_1+2c_2\geq 1$. \qed

\section{Proof of Proposition \ref{pro_wifi_bundle}}\label{app_pro_wifi_bundle}

We divide our bundled pricing analysis into two cases: $p_{12}<\frac{3}{4}$ and $p_{12}\geq\frac{3}{4}$.

If $p_{12}<\frac{3}{4}$, (\ref{equ_n_b*}) doesn't have any real root and $D_{12}<\frac{1-p_{12}}{1-D_{12}}$ always holds, which shows that, for any initial $D_{12}$ value, the number of users subscribing to the service will increase until $D_{12}^*=1$ eventually. That is to say, all users will subscribe to the service. As the profit $\Pi_b(p_{12})=p_{12}-c_1-c_2$ increases with $p_{12}$, the provider would set maximum $p_{12}^*\rightarrow\frac{3}{4}$. This also tells the optimal bundle price should satisfy $p_{12}\geq\frac{3}{4}$.

If $p_{12}\geq \frac{3}{4}$, there are two equilibria $D_{12}^*(p_{12})$ by solving (\ref{equ_n_b*}), i.e., \been  D_{12}^*(p_{12})=\frac{1\pm\sqrt{4p_{12}-3}}{2}. \enen
In the following, we analyze the provider's optimal bundled pricing design under these two equilibria one by one.

\begin{itemize}
  \item When $D_{12}^*(p_{12})=\frac{1-\sqrt{4p_{12}-3}}{2}$, we can verify that bundle demand $\frac{1-p_{12}}{1-D_{12}^*(p_{12})}\leq 1$ is satisfied due to $p_{12}\geq \frac{3}{4}$. Then, the profit is a concave function of $p_{12}$, and is given by  \bee\label{equ_proofpib} \Pi_b(p_{12})=\frac{1-\sqrt{4p_{12}-3}}{2}(p_{12}-c_1-c_2), \ene The optimal bundle price $p_{12}^*$ can be derived by solving the first-order condition:
\bee\label{equ_proofderive} 36p_{12}^2-(24(c_1+c_2)+40)p_{12}+(2(c_1+c_2)+3)^2+3=0. \ene
If (\ref{equ_proofderive}) has real solution, we can obtain the optimal profit $\Pi_b(p_{12}^*)$ by substituting $p_{12}^*$ into (\ref{equ_proofpib}). If (\ref{equ_proofderive}) doesn't have a real solution, the left-hand-side of (\ref{equ_proofderive}) is always positive, which results in $\frac{\partial \Pi_b}{\partial p_{12}}<0$. Thus, $\Pi_b(p_{12})$ decreases with $p_{12}$ and $p_{12}^*=\frac{3}{4}$.
  \item When $D_{12}^*(p_{12})=\frac{1+\sqrt{4p_{12}-3}}{2}$, we also have bundle demand $\frac{1-p_{12}}{1-D_{12}^*(p_{12})} \leq 1$ due to $p_{12}\geq \frac{3}{4}$. Since the profit \been \Pi_b(p_{12})=\frac{1+\sqrt{4p_{12}-3}}{2}(p_{12}-c_1-c_2) \enen increases with $p_{12}$, the provider would set $p_{12}^*=1$ to reach $D_{12}^*=1$ and maximize its profit.
\end{itemize}

By comparing the optimal profits received from these two demand equilibria, the provider chooses the bundle price $p_{12}^*=1$ for reaching the maximum demand $D_{12}^*=1$ and the maximum profit $1-c_1-c_2$.

\section{Proof of Proposition \ref{pro_sep_mb}}\label{app_pro_sep_mb}

According to (\ref{equ_sep_mb_D1}) and (\ref{equ_sep_mb_D3}), to determine the equilibrium demands $D_1^*$ and $D_{12}^*$, we need to compare $\frac{p_{12}-p_1}{D_1+D_{12}}$ and $1-p_1$, $\frac{1-p_{12}}{1-D_1-D_{12}}$ and $1$.
%
%, there are many cases based on the relationships between $\frac{1-p_{12}}{1-D_1-D_{12}}$ and $1$, $\frac{p_{12}-p_1}{D_1+D_{12}}$ and $1-p_1$. It is trivial to analyze all these cases. In the following, we show the case when $\Pi_h(p_1, p_{12})$ achieves its maximum profit.

Consider the case when $\frac{1-p_{12}}{1-D_1-D_{12}}\geq 1$ and $\frac{p_{12}-p_1}{D_1+D_{12}}\leq 1-p_1$, by solving (\ref{equ_sep_mb_D1}) and (\ref{equ_sep_mb_D3}) jointly, the equilibrium device demand $D_1^*$ is
\been D_1^*=p_{12}-p_1, \enen
and the equilibrium bundle demand $D_{12}^*$ is
\been D_{12}^*=1-p_{12}+p_1. \enen

The total demand is $D_1^*+D_{12}^*=1$ and the provider would like to set the
hybrid pricing such that all users are involved in the market. Then, the objective of the provider is to maximize

\been\begin{split} \Pi_h(p_1,p_{12})=&(p_1-c_1)(p_{12}-p_1)\\
&+(p_{12}-c_1-c_2)(1-p_{12}+p_1)\\
=&p_{12}-c_1-c_2+(p_{12}-p_1)(c_2-p_{12}+p_1). \end{split}\enen

By taking derivative of $\Pi_h(p_1,p_{12})$ over $p_1$, we get the relationship between optimal $p_{12}^*$ and $p_{1}^*$ as \been p_{12}^*=p_1^*+\frac{c_2}{2}. \enen

By substituting it back to the profit $\Pi_h(p_1,p_{12})$, we can derive $p_1^*=1-\frac{c_2}{2}$ and $p_{12}^*=1$. The corresponding demands are $D_1^*=\frac{c_2}{2}$ and $D_{12}^*=1-\frac{c_2}{2}$. By substituting $p_{12}^*$ and $p_{1}^*$ to $\Pi_h(p_1,p_{12})$, we obtain the optimal profit in (\ref{equ_sep_mb_opR}).

%Since $p_{12}^*=1$ is the maximum bundle price due to the constraint $p_{12}\leq 1$.

For all other three cases, i.e., ($\frac{1-p_{12}}{1-D_1-D_{12}}\leq 1$, $\frac{p_{12}-p_1}{D_1+D_{12}}\leq 1-p_1$), or ($\frac{1-p_{12}}{1-D_1-D_{12}}\geq 1$, $\frac{p_{12}-p_1}{D_1+D_{12}}\geq 1-p_1$), or ($\frac{1-p_{12}}{1-D_1-D_{12}}\leq 1$, $\frac{p_{12}-p_1}{D_1+D_{12}}\geq 1-p_1$), analysis follows similarly and we can verify that the optimal prices should be set such that $\frac{1-p_{12}^*}{1-D_1^*-D_{12}^*}=1$ and $\frac{p_{12}^*-p_1^*}{D_1^*+D_{12}^*}=1-p_1^*$ to involve all users in the market. Thus, all these three cases converge to the first case and the optimal profit is (\ref{equ_sep_mb_opR}) with the optimal prices $p_1^*=1-\frac{c_2}{2}, p_{12}^*=1$.

\section{Proof of Proposition \ref{pro_bundle}}\label{app_pro_bundle}

In medium price regime ($1+\lambda D_{12}<p_{12}<\bt+\lambda D_{12}$ or simply $\frac{1}{2\bt}\leq D_{12}\leq 1-\frac{1}{2\bt}$), the users' subscription and their demand are shown in Fig. \ref{pure2}. By calculating the shared trapezoid area, the equilibrium demand can be derived from
\bee\begin{split} D_{12}=\frac{(\bt-(p_{12}-\lambda D_{12}))+(\bt-(p_{12}-1-\lambda D_{12}))}{2\bt}, \end{split}\ene
which provides
\bee D_{12}^*(p_{12})=\frac{2\bt-2p_{12}+1}{2\bt-2\lambda}. \ene

Substitute $D_{12}^*(p_{12})$ to profit $\Pi_b$, we find $\Pi_b(p_{12})$ as a concave function of $p_{12}$. By checking the first-order condition, we obtain the equilibrium price $p_{12}^*$ and the resulting equilibrium demand $D_{12}^*$ as
\bee\label{equ_p12(ii)} p_{12}^*=\frac{2\bt+2(c_1+c_2)+1}{4}, \ene
\bee\label{equ_content_iiD} D_{12}^*=\frac{\bt-c_1-c_2+\frac{1}{2}}{2(\bt-\lambda)}. \ene
The demand should still satisfy our initial assumption $\frac{1}{2\bt}\leq D_{12}^*\leq 1-\frac{1}{2\bt}$ in the medium price regime. That requires the following two conditions:
\bee 2\bt (c_1+c_2)+\bt\leq2(\bt^2+\lambda), \ene
and \bee 2(\bt^2+\bt (c_1+c_2)+\lambda)\geq 3\bt+4\bt\lambda. \ene

In low price regime ($\lambda D_{12}<p_{12}<1+\lambda D_{12}$ or simply $D_{12}>1-\frac{1}{2\bt}$), the users' subscription and their demand are shown in Fig. \ref{pure3}. Calculate the shared area by subtracting the normalized equal-right triangle's area (of side length $p_{12}-\lambda D_{12}$) from $1$, the equilibrium demand can be derived from
\bee\label{equ_d12(iii)} D_{12}=1-\frac{(p_{12}-\lambda D_{12})^2}{2\bt}. \ene
%i.e., $p_{12}<1+\lambda D_{12}$,

From (\ref{equ_d12(iii)}), we can show $p_{12}$ as a unique function of demand $D_{12}$, that is
\bee\label{equ_p12(iii)} p_{12}^*(D_{12})=\sqrt{2\bt(1-D_{12})}+\lambda D_{12}. \ene
Substitute it to profit $\Pi_b$, the provider's optimization problem in Stage I is
\been \max_{D_{12}}\Pi_b(D_{12})=(\sqrt{2\bt(1-D_{12})}+\lambda D_{12}-c_1-c_2)D_{12}. \enen

By checking the first-order condition, we have the equilibrium demand $D_{12}^*$ as the solution to
\bee\begin{split}\label{equ_content_iiiD} &2\sqrt{2\bt}\lambda(1-D_{12})^{\frac{3}{2}}-(2\lambda-c_1-c_2)\sqrt{2\bt}\sqrt{1-D_{12}}\\
  &-3\bt(1-D_{12})+\bt=0. \end{split}\ene

We can show that there is a unique solution to (\ref{equ_content_iiiD}), satisfying assumption $D_{12}^*\in(1-\frac{1}{2\bt},1]$ under the following condition:
\bee 2(\bt^2+\bt (c_1+c_2)+\lambda)<3\bt+4\bt\lambda.\ene

\section{Proof of Proposition \ref{pro_4.1}}\label{app_pro_4.1}

As explained for hybrid pricing in Section \ref{sec_content_hybrid}, we require that constraint (\ref{equ_condition_p1_2}) holds for non-zero device-only demand. Additionally, $p_{12}-p_1\leq 1+\lambda D_{12}$ should hold for non-zero bundle demand. Thus, we have $\lambda D_{12}\leq p_{12}-p_1\leq 1+\lambda D_{12}$. Given this feasible range of $p_{12}-p_1$, we introduce a new decision variable $\sigma\in[0,1]$ such that $p_{12}=p_1+\sigma+\lambda D_{12}$. We can also rewrite device-only demand in (\ref{equ_D1sep}) as $D_1^*=\frac{(\bar{\theta}-p_1)\sigma}{\bt}$,
where $p_1\in[0,\bar{\theta}]$ and $D_1^*$ is independent of $\lambda$.
Provided with $p_{12}=p_1+\sigma+\lambda D_{12}^*$, the provider's profit under hybrid pricing is
\bee\label{equ_pimcompare} \Pi_h=(p_1+\sigma+\lambda D_{12}-c_1-c_2)D_{12}+(p_1-c_1)D_1. \ene

Note that $p_1\leq \bt$, we have $\frac{\frac{\partial \Pi_h}{\partial\sigma}}{\partial \lambda}<0$ for the two cases $p_{12}\leq 1+\lambda D_{12}$ and $p_{12}>1+\lambda D_{12}$ shown in Fig. \ref{hybridfig}. Thus, $\frac{\partial \Pi_h}{\partial\sigma}$ decreases with $\lambda$, which means for any given $c_1, c_2$ and $\bt$, $\frac{\partial \Pi_h}{\partial\sigma}$ changes from positive to negative as $\lambda$ increases. As $\sigma$ decreases, hybrid pricing degenerates to bundled pricing (see Fig. \ref{hybridfig}). Thus, when $\frac{\partial \Pi_h}{\partial\sigma}$ is negative, the provider receives the optimal profit under bundled pricing. Therefore, we can conclude that the provider is more likely to sell the device and service as a bundle as $\lambda$ increases.

In the following, we will discuss the condition when the bundled pricing outperforms the hybrid pricing for the two cases shown in Fig. \ref{hybridfig}.%Note that the hybrid pricing degenerates to bundled pricing when $D_1^*=0$.

When $p_{12}\leq 1+\lambda D_{12}$, i.e., $\sigma+p_1\leq 1$, the equilibrium bundle demand is
\begin{equation*} D_{12}^*=\frac{\bt(1-\sigma)-\frac{p_1^2}{2}}{\bt}. \end{equation*}

Substitute $D_1^*$ and $D_{12}^*$ into $\Pi_h$, we have $\frac{\partial \Pi_h}{\partial\sigma}$ as a concave function of $p_1$. Insert the optimal $p_1^*$ into $\frac{\partial \Pi_h}{\partial\sigma}$, we can check that $\frac{\partial \Pi_h(p_1^*)}{\partial\sigma}<0$ when $\lambda>\frac{3}{2}$ and $c_2<1$. Therefore, the provider's profit $\Pi_h$ reaches its maximum when $\sigma=0$, which means $D_1^*=0$. That is, if $\lambda>\frac{3}{2}$ and $c_2<1$, the hybrid pricing degenerates to bundled pricing when $p_{12}\leq 1+\lambda D_{12}$ (see Fig. \ref{mixed1}).

When $p_{12}>1+\lambda D_{12}$, i.e., $\sigma+p_1>1$, the equilibrium bundle demand is
\begin{equation*} D_{12}^*=\frac{(2\bt-2p_1+1-\sigma)(1-\sigma)}{2\bt}. \end{equation*}

Similarly, we can check that $\frac{\partial \Pi_h}{\partial\sigma}$ is concave with $p_1$. Insert the optimal $p_1^*$ into $\frac{\partial \Pi_h}{\partial\sigma}$ and then we have $\frac{\partial \Pi_h(p_1^*)}{\partial\sigma}<0$ always holds when $4c_1+c_2<1$. That is, smaller $\sigma$ leads to larger profit. As shown in Fig. \ref{mixed2}, when $p_{12}-p_1$ decreases until $\sigma=0$ if $p_{12}\leq \bt+\lambda D_{12}$ or $p_1=\bt$ if $p_{12}>\bt+\lambda D_{12}$, the hybrid pricing degenerates to bundled pricing.

In conclusion, when $\lambda>\frac{3}{2}$ and $4c_1+c_2<1$, the hybrid pricing degenerates to bundled pricing.

\section{Proof of Proposition \ref{pro_hybrid_connectivity}}\label{app_pro_hybrid_connectivity}

Note that hybrid pricing degenerates to bundled pricing if the optimal device price $p_1^*=p_{12}$ for any given bundled price $p_{12}$. This tells that device-only price is unreasonably high and no user will choose it, as bundle option at the same price provides higher utility. We only need to prove that for any given $p_{12}$, the provider will not set $p_1$ equal to $p_{12}$, or simply $\frac{\partial \Pi_h}{\partial p_1}|_{p_1=p_{12}}<0$. For any continuous distribution of $\alpha$ with pdf  $f_{\alpha}$, Lemma \ref{lem_mixbundle} still holds. Under hybrid pricing, the equilibrium demand for device only is $D_1^*=\int_0^{p_{12}-p_1}f_{\alpha}(\alpha)d\alpha$ and the equilibrium demand for bundled product is $D_{12}^*=\int_{p_{12}-p_1}^1f_{\alpha}(\alpha)d\alpha$. By substituting such equilibrium demands into $\Pi_h$ in (\ref{equ_Pimixbundle}), we have $\frac{\partial \Pi_h}{\partial p_1}|_{p_1=p_{12}}=-c_2f_{\alpha}(0)<0$. Thus, the provider will lower the device-only price and always choose hybrid pricing rather than bundled pricing.

\bibliographystyle{IEEEtran}
\bibliography{sigproc}

\begin{thebibliography}{10}
\providecommand{\url}[1]{#1}
\csname url@samestyle\endcsname
\providecommand{\newblock}{\relax}
\providecommand{\bibinfo}[2]{#2}
\providecommand{\BIBentrySTDinterwordspacing}{\spaceskip=0pt\relax}
\providecommand{\BIBentryALTinterwordstretchfactor}{4}
\providecommand{\BIBentryALTinterwordspacing}{\spaceskip=\fontdimen2\font plus
\BIBentryALTinterwordstretchfactor\fontdimen3\font minus
  \fontdimen4\font\relax}
\providecommand{\BIBforeignlanguage}[2]{{%
\expandafter\ifx\csname l@#1\endcsname\relax
\typeout{** WARNING: IEEEtran.bst: No hyphenation pattern has been}%
\typeout{** loaded for the language `#1'. Using the pattern for}%
\typeout{** the default language instead.}%
\else
\language=\csname l@#1\endcsname
\fi
#2}}
\providecommand{\BIBdecl}{\relax}
\BIBdecl

\bibitem{fon35}
INFORMILO, \emph{``Top 25 global mobile start-ups," [Online]. Available: {\rm
  http://www.informilo.com/2014/02/top-25-global-mobile-start-ups-2/}}.

\bibitem{iwatch2016}
D.~Wakabayashi, ``Apple's watch outpaced the iphone in first year,'' \emph{The
  Wall Street Journal}, 2016.

\bibitem{consumerlab10hot}
E.~Consumerlab, ``10 hot consumer trends 2016,'' \emph{Stockholm: Ericsson},
  2016.

\bibitem{niyato2007wireless}
D.~Niyato and E.~Hossain, ``Wireless broadband access: Wimax and
  beyond-integration of wimax and wifi: Optimal pricing for bandwidth
  sharing,'' \emph{IEEE Communications Magazine}, vol.~45, no.~5, pp. 140--146,
  2007.

\bibitem{afrasiabi2012pricing}
M.~H. Afrasiabi and R.~Gu{\'e}rin, ``Pricing strategies for user-provided
  connectivity services,'' in \emph{Proceedings of IEEE INFOCOM}, 2012, pp.
  2766--2770.

\bibitem{openspark}
OpenSpark, \emph{{\rm https://open.sparknet.fi/index.php}}.

\bibitem{whisher}
Whisher, \emph{{\rm https://whisher.jaleco.com/}}.

\bibitem{rafiei2013product}
H.~Rafiei, M.~Rabbani, J.~Razmi, and F.~Jolai, ``Product bundle pricing in the
  new millennium: A literature review,'' \emph{International Journal of
  Advances in Management Science}, 2013.

\bibitem{lee2016analysis}
S.~Lee, S.~Nam, and H.~Shin, ``The analysis of sleep stages with motion and
  heart rate signals from a handheld wearable device,'' in \emph{International
  Conference on Information and Communication Technology Convergence
  (ICTC)}.\hskip 1em plus 0.5em minus 0.4em\relax IEEE, 2016, pp. 1135--1137.

\bibitem{gizelis2011survey}
C.~A. Gizelis and D.~D. Vergados, ``A survey of pricing schemes in wireless
  networks,'' \emph{IEEE Communications Surveys \& Tutorials}, vol.~13, no.~1,
  pp. 126--145, 2011.

\bibitem{duan2013economics}
L.~Duan, J.~Huang, and B.~Shou, ``Economics of femtocell service provision,''
  \emph{IEEE Transactions on Mobile Computing}, vol.~12, no.~11, pp.
  2261--2273, 2013.

\bibitem{feldman2013pricing}
M.~Feldman, D.~Kempe, B.~Lucier, and R.~Paes~Leme, ``Pricing public goods for
  private sale,'' in \emph{Proceedings of the fourteenth ACM conference on
  Electronic commerce}.\hskip 1em plus 0.5em minus 0.4em\relax ACM, 2013, pp.
  417--434.

\bibitem{munagala2014value}
K.~Munagala and X.~Xu, ``Value-based network externalities and optimal auction
  design,'' in \emph{International Conference on Web and Internet
  Economics}.\hskip 1em plus 0.5em minus 0.4em\relax Springer, 2014, pp.
  147--160.

\bibitem{prasad2010optimal}
A.~Prasad, R.~Venkatesh, and V.~Mahajan, ``Optimal bundling of technological
  products with network externality,'' \emph{Management Science}, vol.~56,
  no.~12, pp. 2224--2236, 2010.

\bibitem{yan2011profit}
R.~Yan and S.~Bandyopadhyay, ``The profit benefits of bundle pricing of
  complementary products,'' \emph{Journal of Retailing and Consumer Services},
  vol.~18, no.~4, pp. 355--361, 2011.

\bibitem{cohen2011truth}
E.~Cohen, M.~Feldman, A.~Fiat, H.~Kaplan, and S.~Olonetsky, ``Truth, envy, and
  truthful market clearing bundle pricing,'' in \emph{International Workshop on
  Internet and Network Economics}.\hskip 1em plus 0.5em minus 0.4em\relax
  Springer, 2011, pp. 97--108.

\bibitem{wu2014exploring}
W.~Wu, R.~T. Ma, and J.~C. Lui, ``Exploring bundling sale strategy in online
  service markets with network effects,'' in \emph{INFOCOM, 2014 Proceedings
  IEEE}.\hskip 1em plus 0.5em minus 0.4em\relax IEEE, 2014, pp. 442--450.

\bibitem{gong2016jsac}
X.~Gong, L.~Duan, X.~Chen, and J.~Zhang, ``When social network effect meets
  congestion effect in wireless networks: Data usage equilibrium and optimal
  pricing,'' \emph{IEEE Journal on Selected Areas in Communications}, vol.~35,
  no.~2, pp. 449--462, 2017.

\bibitem{niyato2008competitive}
D.~Niyato and E.~Hossain, ``Competitive pricing for spectrum sharing in
  cognitive radio networks: Dynamic game, inefficiency of nash equilibrium, and
  collusion,'' \emph{IEEE Journal on selected areas in communications},
  vol.~26, no.~1, 2008.

\bibitem{duan2012duopoly}
L.~Duan, J.~Huang, and B.~Shou, ``Duopoly competition in dynamic spectrum
  leasing and pricing,'' \emph{IEEE Transactions on Mobile Computing}, vol.~11,
  no.~11, pp. 1706--1719, 2012.

\bibitem{duan2015pricing}
------, ``Pricing for local and global wi-fi markets,'' \emph{IEEE Transactions
  on Mobile Computing}, vol.~14, no.~5, pp. 1056--1070, 2015.

\bibitem{manshaei2008wireless}
M.~H. Manshaei, J.~Freudiger, M.~F{\'e}legyh{\'a}zi, P.~Marbach, and J.-P.
  Hubaux, ``Wireless social community networks: a game-theoretic analysis,'' in
  \emph{Communications, 2008 IEEE International Zurich Seminar on}.\hskip 1em
  plus 0.5em minus 0.4em\relax IEEE, 2008, pp. 22--25.

\bibitem{enck2011defending}
W.~Enck, ``Defending users against smartphone apps: Techniques and future
  directions,'' in \emph{International Conference on Information Systems
  Security}.

\end{thebibliography}

\begin{IEEEbiography}
[{\includegraphics[width=1in,height=1.25in,clip,keepaspectratio]{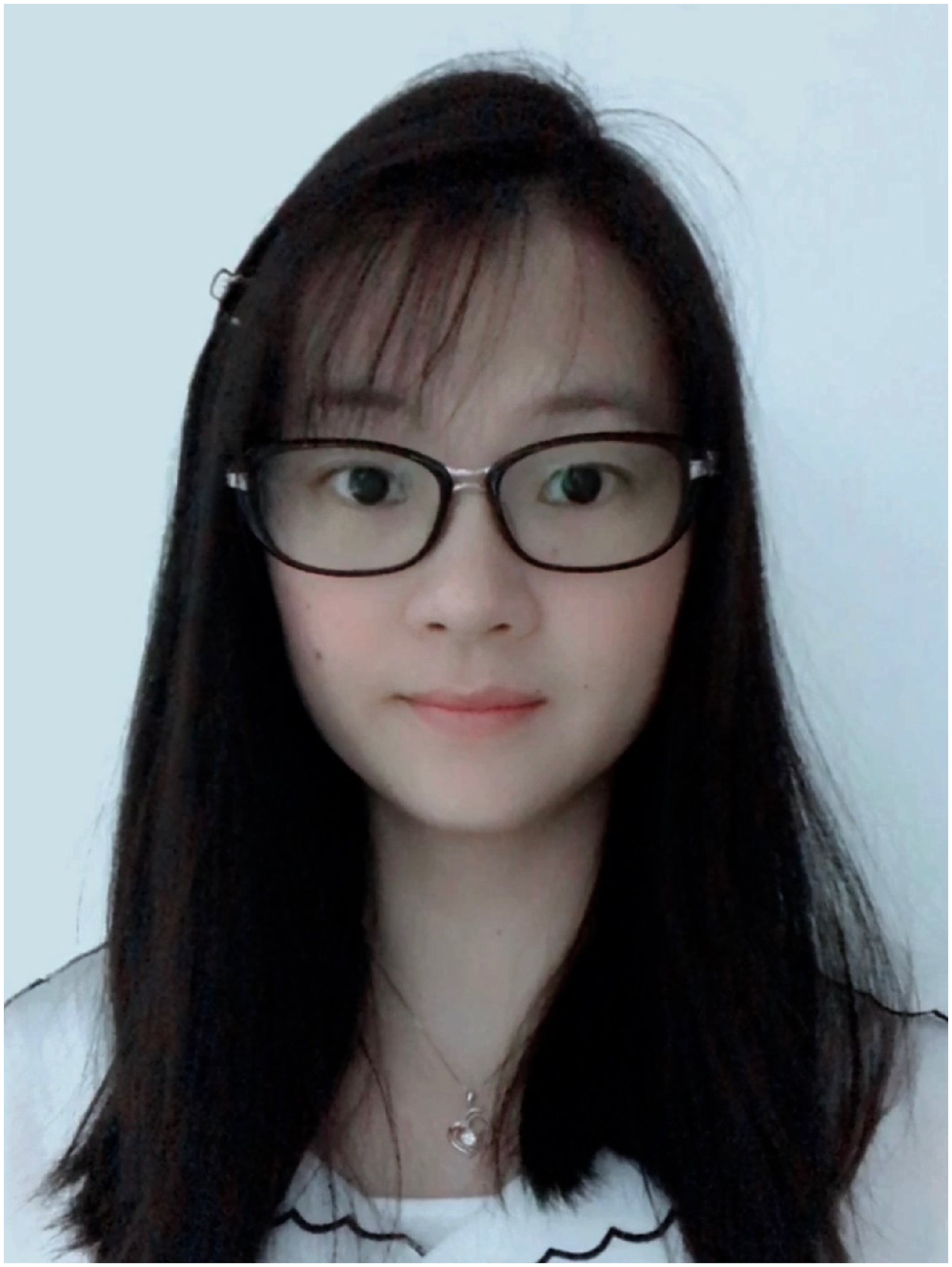}}]{Xuehe Wang} (S'15-M'16)
received her bachelor
degree in mathematics from Sun Yat-sen
University, China in 2011, and her Ph.D. degree in electrical and electronic engineering from Nanyang Technological University, Singapore in 2016. She is a postdoctoral research fellow with the Pillar of Engineering Systems and Design, Singapore University of Technology and Design. Her
research interest covers game theory, cooperative control theory, and network economics.
\end{IEEEbiography}

\begin{IEEEbiography}
[{\includegraphics[width=1in,height=1.25in,clip,keepaspectratio]{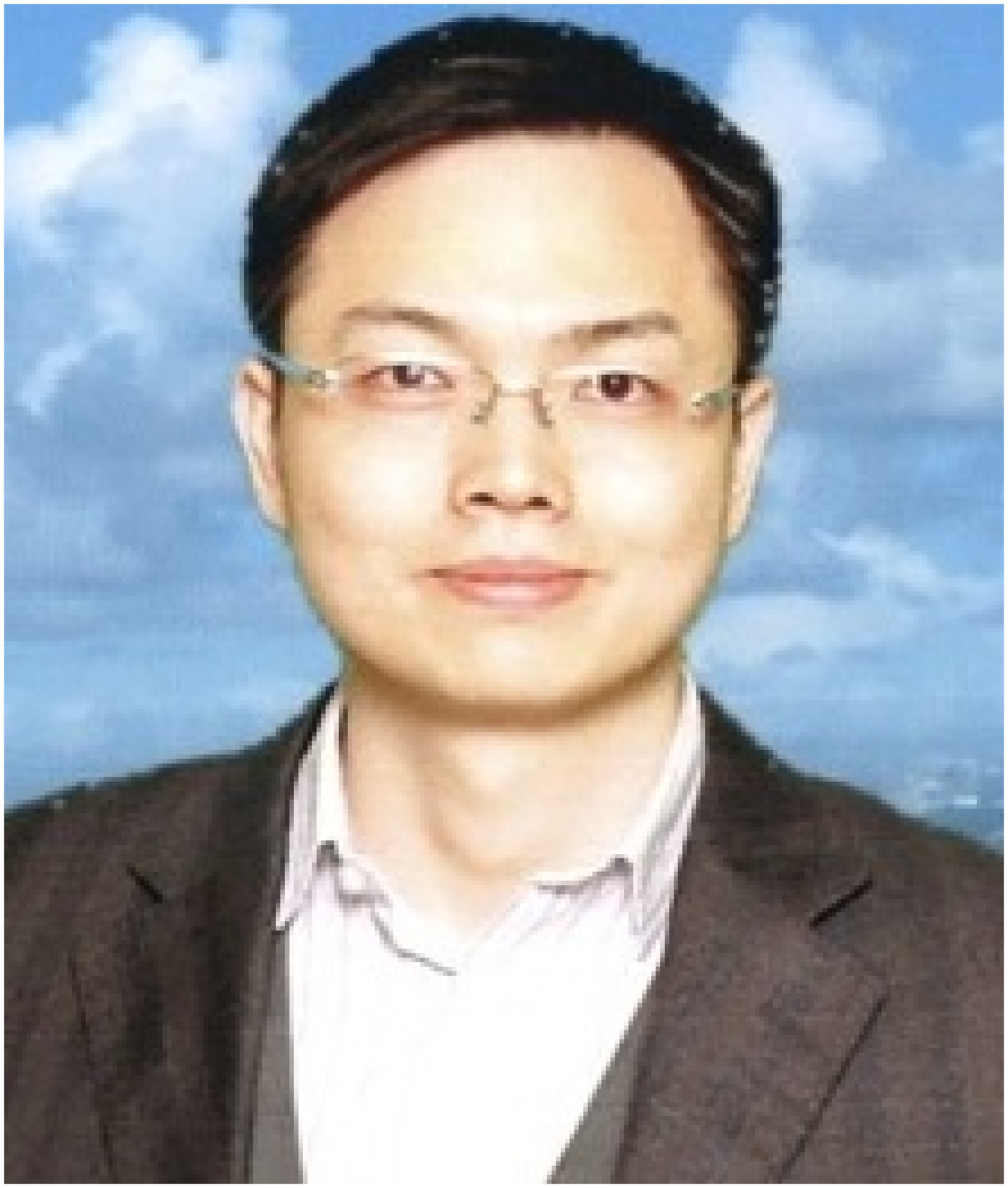}}]{Lingjie Duan} (S'09-M'12) received the B.Eng.
degree from the Harbin Institute of Technology
in 2008, and the Ph.D. degree from The Chinese
University of Hong Kong in 2012. He is an Assistant
Professor of Engineering Systems and Design
with the Singapore University of Technology and
Design (SUTD). In 2011, he was a Visiting Scholar
with the Department of Electrical Engineering and
Computer Sciences, University of California at
Berkeley, Berkeley, CA, USA. He is currently leading
the Network Economics and Optimization Laboratory,
SUTD. His research interests include network economics and game
theory, cognitive communications and cooperative networking, energy harvesting
wireless communications, and network security. He is an Editor of
the IEEE COMMUNICATIONS SURVEYS AND TUTORIALS, and is a SWAT
Team Member of the IEEE TRANSACTIONS ON VEHICULAR TECHNOLOGY.
He currently serves as a Guest Editor of the IEEE JOURNAL ON SELECTED
AREAS IN COMMUNICATIONS Special Issue on Human-in-the-Loop Mobile
Networks, and also serves as a Guest Editor of the IEEE Wireless Communications
Magazine. He also serves as a Technical Program Committee Member
of many leading conferences in communications and networking (e.g.,
ACM MobiHoc, IEEE SECON, INFOCOM, WiOPT, and NetEcon). He received the
10th IEEE ComSoc Asia-Pacific Outstanding Young Researcher Award in
2015 and the Hong Kong Young Scientist Award (Finalist in Engineering Science
track) in 2014.
\end{IEEEbiography}

\begin{IEEEbiography}
[{\includegraphics[width=1in,height=1.25in,clip,keepaspectratio]{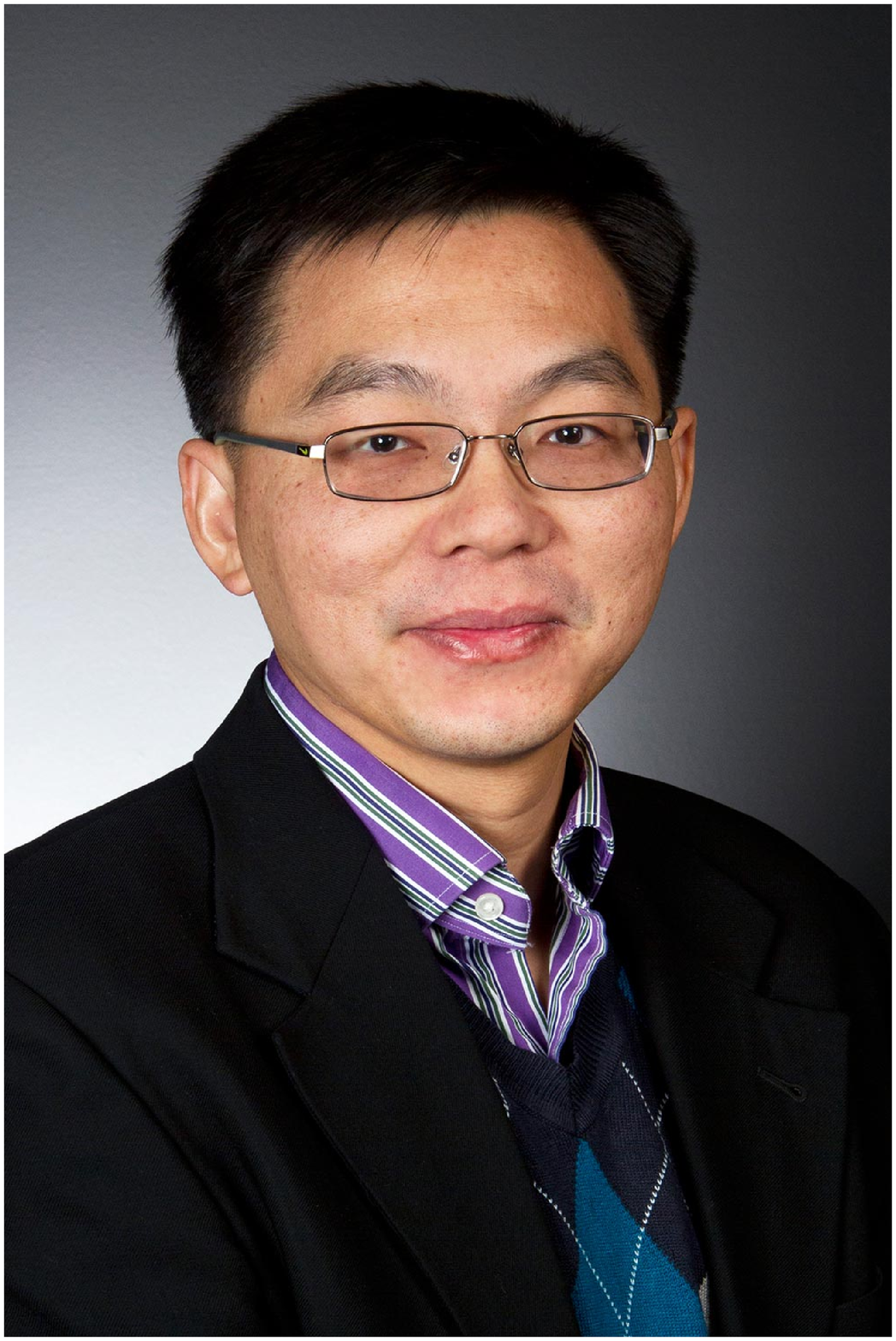}}] {Junshan Zhang} (S'98-M'00-SM'06-F'12) received his Ph.D. degree from the School of ECE at Purdue University in 2000. He joined the School of ECEE at Arizona State University in August 2000, where he has been Fulton Chair Professor since 2015. His research interests fall in the general field of information networks and its intersections with social networks and power networks. His current research focuses on fundamental problems in information networks and energy networks, including Fog Computing and its applications in IoT and 5G, optimization/control of mobile social networks and cognitive radio networks, modeling and optimization for smart grid, and privacy/security in information networks.

Prof. Zhang is a fellow of the IEEE, and a recipient of the ONR Young Investigator Award in 2005 and the NSF CAREER award in 2003. He received the IEEE Wireless Communication Technical Committee Recognition Award in 2016. His papers have won a few awards, including the Kenneth C. Sevcik Outstanding Student Paper Award of ACM SIGMETRICS/IFIP Performance 2016, the Best Paper Runner-up Award of IEEE INFOCOM 2009 and IEEE INFOCOM 2014, and the Best Paper Award at IEEE ICC 2008. He was TPC co-chair for a number of major conferences in communication networks, including IEEE INFOCOM 2012 and ACM MOBIHOC 2015. He was the general chair for WiOpt 2016 and IEEE Communication Theory Workshop 2007. He was an Associate Editor for IEEE Transactions on Wireless Communications, an editor for the Computer Network journal and an editor IEEE Wireless Communication Magazine. He was a Distinguished Lecturer of the IEEE Communications Society. He is currently serving as an editor-at-large for IEEE/ACM Transactions on Networking and an editor for IEEE Network Magazine.
\end{IEEEbiography}

%% Can use something like this to put references on a page
%% by themselves when using endfloat and the captionsoff option.
%\ifCLASSOPTIONcaptionsoff
%  \newpage
%\fi
%

%\end{IEEEbiography}
%% insert where needed to balance the two columns on the last page with
%%\newpage


%\vfill


\end{document}